\documentclass[aps.pra,twocolumn]{revtex4}

\usepackage{graphicx,xcolor}
\usepackage{times}
\usepackage{amsmath,amssymb}
\usepackage[normalem]{ulem}
\usepackage[colorlinks,citecolor=red]{hyperref}

\newcommand\bra[1]{\mathinner{\langle{\textstyle#1}\rvert}}
\newcommand\ket[1]{\mathinner{\lvert{\textstyle#1}\rangle}}
\newcommand{\Braket}[1]{\mathinner{\left\langle{#1}\right\rangle}}

\newcommand{\bfS}{{\mathbf{S}}}
\newcommand{\bfs}{{\mathbf{s}}}
\newcommand{\bftau}{{\boldsymbol{\tau}}}

\newcommand{\im}{\mathrm{Im}}

\let\up=\uparrow
\let\down=\downarrow
\let\Up=\Uparrow
\let\Down=\Downarrow

\newcommand{\eqnref}[1]{Eq.~(\ref{#1})}
\newcommand{\eqnsref}[1]{Eqs.~(\ref{#1})}

\newcommand{\figref}[1]{Fig.~\ref{#1}}
\newcommand{\figsref}[1]{Figs.~\ref{#1}}
\newcommand{\Figref}[1]{Figure~\ref{#1}}

\newcommand{\secref}[1]{Sec.~\ref{#1}}

\begin{document}

\title[Mixed-Valence Transition on a Quantum-Dot System]{Mixed-Valence
  Transition on a Quantum-Dot Coupled to Superconducting and Spin-Polarized
  Leads}

\author{Minchul Lee}
\address{Department of Applied Physics and Institute of Natural Science, College of Applied Science, Kyung Hee University, Yongin 17104, Korea}

\author{Mahn-Soo Choi}

\email{choims@korea.ac.kr}

\address{Department of Physics, Korea University, Seoul 02841, Korea}

\begin{abstract}
We consider a quantum dot coupled to both superconducting and spin-polarized electrodes, and study the triad interplay of the Kondo effect, superconductivity, and ferromagnetism, any pair of which compete with and suppress each other.
We find that the interplay leads to a mixed-valence quantum phase transition,
which for other typical sysmstems is merely a crossover rather than a true transition.
At the transition, the system changes from the spin doublet to singlet state. The singlet phase is adiabatically connected (through crossovers) to the so-called 'charge Kondo state' and to the superconducting state. We analyze in detail the physical characteristics of different states and propose that the
measurement of the cross-current correlation and the charge relaxation
resistance can clearly distinguish between them.
\end{abstract}


\maketitle

\section{Introduction}
\label{sec:introduction}

Superconductivity, ferromagnetism, and Kondo effect are the representative correlation effects in condensed matter physics. Interestingly, any pair of these three effects compete with each other:
Hampering the spin-singlet pairing in ($s$-wave) superconductors, ferromagnetism naturally suppresses superconductivity.
Kondo effect is attributed to another kind of spin-singlet correlation between the itinerant spins in the conduction band and the localized spin on the quantum dot (or magnetic impurity), and hence is suppressed in the presence of ferromagnetism in the conduction band \cite{Lopez02a,Fiete02a,Martinek03b,Martinek03a,ChoiMS04a,Pasupathy04a,Yang11a}.
Energetically, when the exchange Zeeman splitting due to the ferromagnetism is
larger than the Kondo temperature $T_K$ (in the absence of ferromagnetism), the Kondo effect is destroyed.
The competition between the superconducting pairing correlation and the Kondo correlation even leads to a quantum phase transition:
When the superconductivity dominates over the Kondo effect (i.e., the superconducting gap energy $\Delta_0$ larger than the normal-state $T_K$),
the ground states of the system form a doublet owing to the Coulomb blockade on the quantum dot.
In the opposite case ($\Delta_0<T_K$), the quantum dot overcomes the Coulomb blockade and resonantly transports Cooper pairs and the whole system resides in a singlet state. The quantum phase transition is manifested by the $0$-$\pi$ quantum phase transition in nano-structure Josephson junctions consisting of a quantum dot (QD) coupled to two superconducting electrodes \cite{Buitelaar02a,Avishai03a,ChoiMS04c,ChoiMS05d,Siano04a,Siano05a,Campagnano04a,Sellier05a,Cleuziou06a,Buizert07a,Grove-Rasmussen07b,ChoiMS08f,Martin-Rodero11a,Franke11a,Delagrange16a,Delagrange16b,Delagrange15a}.

\begin{figure}[b]
\centering
\includegraphics[height=20mm]{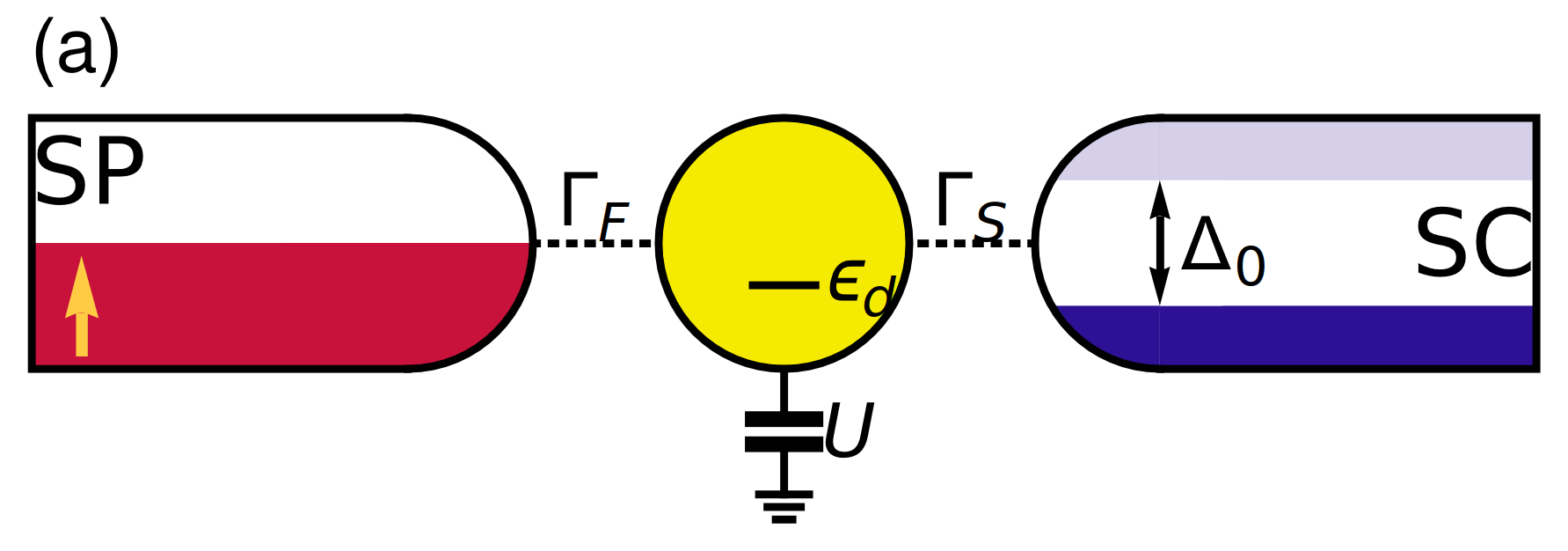}\qquad
\includegraphics[height=20mm]{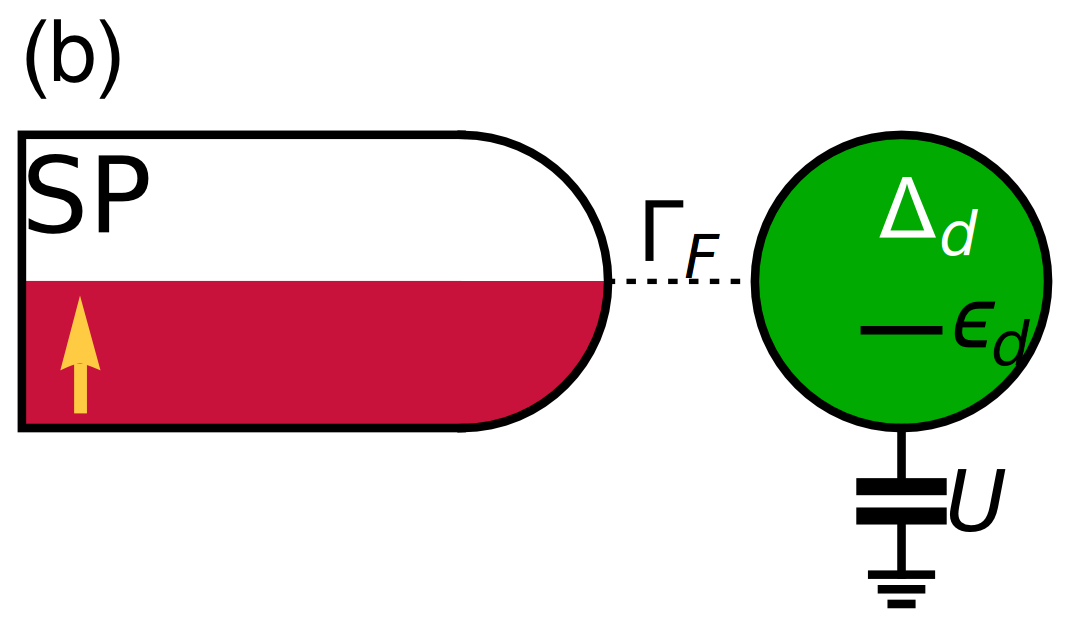}
\caption{System configurations for (a) the spin-polarized (SP) lead-quantum
  dot-superconducting (SC) lead and (b) the spin-polarized lead-quantum dot with the proximity-induced superconductivity. Refer the definition of the symbols to the text.}
\label{fig:system}
\end{figure}

In this work, we study the triad interplay of superconductivity,
ferromagnetism, and Kondo effect all together.
More specifically, we consider a quantum dot coupled to both superconducting
and fully spin-polarized \cite{endnote:2} ferromagnetic electrodes as shown
schematically in \figref{fig:system}~(a).
Similar setups have been studied in different contexts:
exchange-field-dependence of the Andreev reflection \cite{Feng2003jan},
spin-dependent Andreev reflection \cite{Cao2004dec,Weymann15a}, and subgap
states in the QD due to ferromagnetic proximity effect
\cite{Hofstetter2010jun}. The case with a superconducting and two
ferromagnetic leads was also studied to examine the crossed Andreev reflection
\cite{Zhu2001dec,Wojcik14a}.
However, these works either did not properly capture the full correlation effects (that is, Kondo regime could not exploited) \cite{Feng2003jan,Cao2004dec,Zhu2001dec} or studied the modification of Kondo effect due to its interplay with superconductivity and ferromagnetism \cite{Weymann15a,Wojcik14a}.
Note that in the latter works, the Kondo effect survives the relatively weak
superconductivity and/or ferromagnestim.  In this work we explore novel triad
interplays in the opposite limit: Both supercondcutivity and ferromagnestim
are so strong that they \emph{individually} suppress the Kondo effect, but
nevertheless \emph{together} give rise to new resonant transport.

\begin{figure}[t]
\centering
\includegraphics[width=70mm]{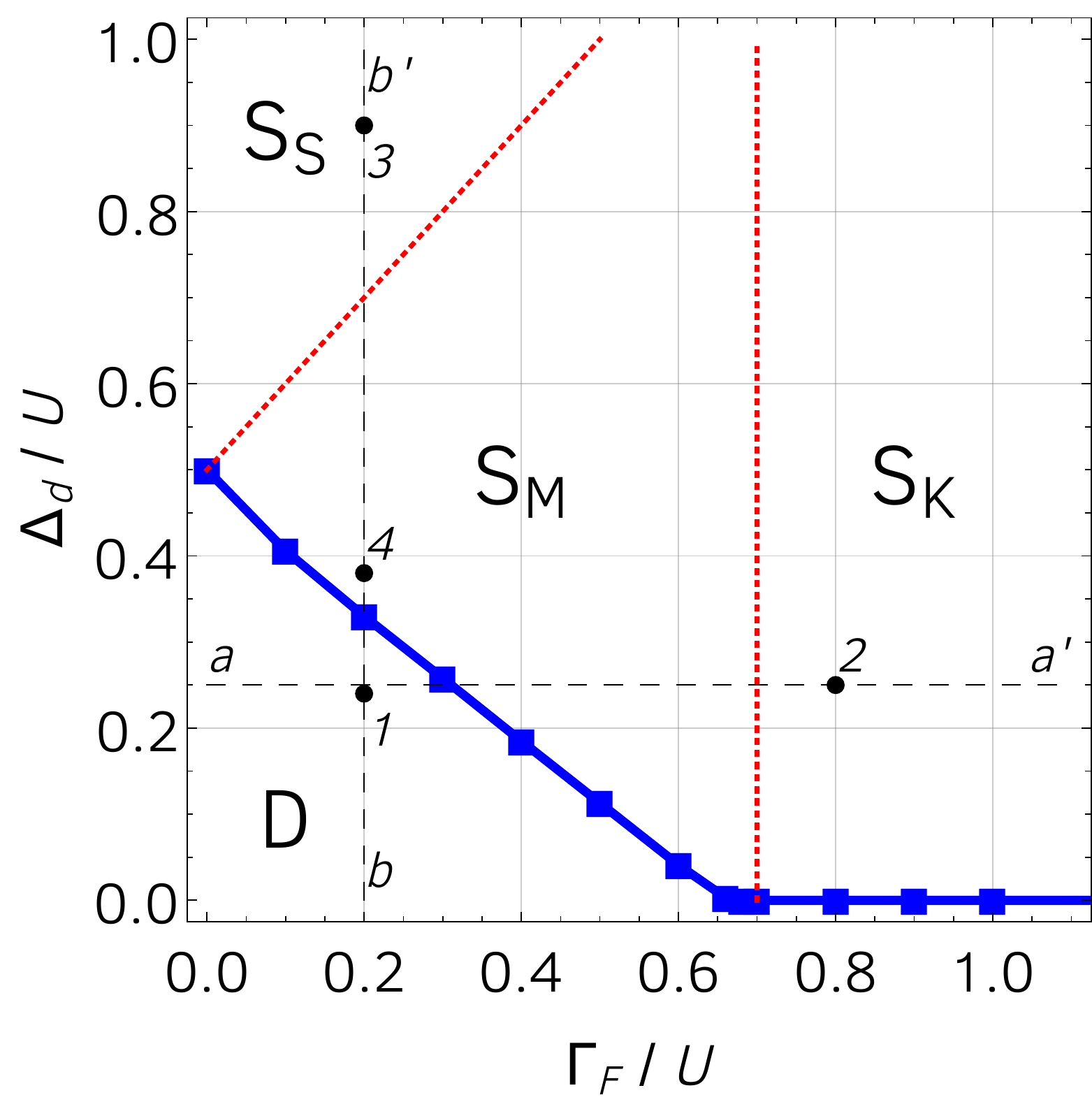}
\caption{(color online) Phase diagram obtained from the NRG method. The phase
  boundary (thick solid line) divides the spin singlet (S) and doublet (D)
  phases. The crossover boundaries (red dotted lines) further divide the
  singlet phase into the superconductivity-dominant (S$_\mathrm{S}$),
  mixed-valence (S$_\mathrm{M}$), and Kondo (S$_\mathrm{K}$) singlet regimes,
  which are connected adiabatically. The black dashed lines are the guides
  along which we examine the change of physical properties of the system.}
\label{fig:pd}
\end{figure}

We find that unlike the aforementioned pairwise competition among the three effects, the triad interplay is ``cooperative'' in certain sense and leads to a new quantum phase transition between doublet and singlet states; see Fig.~\ref{fig:pd}.
The singlet phase is in many respects similar to the mixed-valence state, but
connected adiabatically (through crossovers) to the superconducting state in
the limit of strong coupling to the superconductor and to the `charge Kondo
state' in the limit of strong coupling to the spin-polarized electrode.
The results are obtained with the numerical renormalization group (NRG) method, and the physical explanations are supplemented by other analytic methods such as scaling theory, variational method, and bosonization.
Based on the analysis of the characteristics of the phases, we propose
three experimental methods to identify the phases, which measure the
dot density of state, the cross-current correlation, and the current response
to a small ac gate voltage (charge relaxation resistance), respectively.

The rest of the paper is organized as following:
We describe explicitly our system and the equivalent models for it in Section~\ref{sec:model}. We report our results based on the NRG method, the quantum phase diagram of the system and the characteristic properties of the phases and crossover regions in the singlet phase in Section~\ref{paper::sec:3}. In Section~\ref{sec:discussion}, we apply several analytic methods to provide physical interpretations of the quantum phase transition and the characteristic properties of the different phases and crossover regions. In Section~\ref{sec:experiments}, we discuss possible experiments to observe our findings. Section~\ref{sec:conclusion} summarizes the work and concludes the paper.

\section{Model}
\label{sec:model}

\Figref{fig:system}~(a) shows the schematic configuration of the system of our
interest, in which an interacting quantum dot is coupled to both a
ferromagnetic lead and a superconducting lead. To stress our points, we consider the extreme case where the ferromagnetic lead is fully polarized \cite{endnote:2} and the superconductivity is very strong (the superconducting gap is the largest energy scale). Recall that with the QD coupled to either a fully polarized ferromagnet or a strong superconductor (but not both), neither charge nor spin fluctuations are allowed on the QD.

First highlighting the fully polarized ferromagnetic lead, the Hamiltonian of the system is written as
\begin{equation}
\label{paper::eq:3}
H = H_\mathrm{QD} + H_\mathrm{F} + H_\mathrm{S} + H_\mathrm{T}
\end{equation}
with
\begin{subequations}
\label{paper::eq:4}
\begin{align}
\label{eq:HQD}
H_\mathrm{QD}
& = \delta \sum_\mu (n_\mu - 1/2) + U (n_\up - 1/2) (n_\down - 1/2)
\\
\label{paper::eq:1}
H_\mathrm{F} & = \sum_k \epsilon_k c_{k\up}^\dag c_{k\up}
\\
H_\mathrm{S}
& = \sum_{k\mu} \varepsilon_k a_{k\mu}^\dag a_{k\mu}
- \sum_k (\Delta_0 a_{k\up}^\dag a_{-k\down}^\dag + (h.c.)) \\
H_\mathrm{T}
& = \sum_k (t_F d_\up^\dag c_{k\up} + h.c.)
+ \sum_{k\mu} (t_S d_\mu^\dag a_{k\mu} + h.c.).
\end{align}
\end{subequations}
The operator $d_\mu^\dag$ creates an electron with energy $\epsilon_d$ and spin
$\mu=\up,\down$ and defines the number operator
$n_\mu := d_\mu^\dag d_\mu$; $n_d := \sum_\mu n_\mu$. The dot electrons
interact with each other with the strength $U$.
As mentioned above, the ferromagnetic lead Hamiltonian $H_F$ involves only the majority spin ($\up$) electrons, which are described by the fermion operator $c_{k\up}$ with momentum $k$ and energy $\epsilon_k$.
In the superconducting lead,
the operator $a_{k\mu}$ describes the electron with momenum $k$, spin $\mu$, and single-particle energy $\varepsilon_k$, and the terms in the pairing potential $\Delta_0$ are responsible for the Cooper pairs. Since the
superconducting phase is irrelevant in this study, $\Delta_0$ is assumed to be
real and positive.
The tunnelings between the dot and the ferromagnetic/superconducting leads are
denoted by $t_{F/S}$, respectively, which are assumed to be
momentum-independent for simplicity. The tunnelings induce the hybridizations
$\Gamma_{S/F} := \pi\rho_{S/F}|t_{S/F}|^2$ between the dot and the
superconducting/ferromagnetic leads, respectively, where $\rho_{S/F}$ are the density of states at the Fermi level in the leads.

The parameter $\delta := \epsilon_d + U/2$ indicates the deviation from the particle-hole symmetry.
To make our points clearer and simplify the discussion, in this work we focus
on the particle-hole symmetric case $(\delta=0)$. While the particle-hole
asymmetry gives rise to some additional interesting features~\cite{endnote:1}, the underlying physics can be understood in terms of that in the symmetric case.

Next we exploit the strong superconductivity to further simplify our model: The
pairing gap of the superconducting lead dominates over the other energy scales
($\Delta_0\gg U,\Gamma_S,\Gamma_F$) including $\Delta_0 \gg T_K^0$, where
$T_K^0$ is the Kondo temperature in the absence of ferromagnetic lead $(t_F=0)$
and the superconductivity $(\Delta_0=0)$.  In such a limit, the role of the
superconducting lead is completely manifested in the proximity induced pairing
potential on the QD. Hence, as far as the physics below the superconducting gap
is concerned, the effective low-energy Hamiltonian [see
\figref{fig:system}~(b)] can be approximated, by integrating out the
superconducting degrees of freedom, as
\begin{equation}
\label{eq:H}
H = H_\mathrm{SQD} + H_\mathrm{F} + H_\mathrm{T}
\end{equation}
with
\begin{subequations}
\label{paper::eq:2}
\begin{align}
\label{eq:HSQD}
H_\mathrm{SQD}
&= U \left(n_\up - \frac{1}{2}\right)\left(n_\down - \frac{1}{2}\right)
+ \Delta_d (d_\up^\dag d_\down^\dag + d_\down d_\up) , \\
H_\mathrm{F} & = \sum_k \epsilon_k c_{k\up}^\dag c_{k\up} \,, \\
H_\mathrm{T}
&= \sqrt{\frac{\Gamma_F}{\pi\rho_F}}\sum_k
(d_\up^\dag c_{k\up} + c_{k\up}^\dag d_\up) \,,
\end{align}
\end{subequations}
where the proximity-induced superconducting gap is given by
{$\Delta_d\sim\Gamma_S$} \cite{Volkov95a,McMillan68a}.
In this work, we focus on \eqnref{eq:H} unless specified otherwise.

In passing, the isolated QD with pairing potential \eqref{eq:HSQD} is diagonalized with the eigenstates and the corresponding energies:
\begin{subequations}
\label{eq:iQD}
\begin{align}
\ket{D_\mu^0} & = d_\mu^\dag \ket0,
& E_D^0 & = -U/4 \,,\quad (\mu=\up,\down) \\
\ket{S_\pm^0} & = \frac{1\pm d_\up^\dag d_\down^\dag}{\sqrt{2}}\ket0,
& E_{S\pm}^0 & = U/4\pm \Delta_d
\end{align}
\end{subequations}
The unperturbed ground state of the QD experiences a transition from the spin
doublet state $\ket{D_\mu^0}$ to the spin singlet state $\ket{S_-^0}$ at
$\Delta_d/U=1/2$.

% \subsection{Bogoliubov Transformations}
\subsection{Relation to Other Models}
\label{paper::sec:2.1}

Upon the Bogoliubov-de Gennes (BdG) transformation
\begin{equation}
\label{paper::eq:5}
\begin{bmatrix}
d_\up \\ d_\down^\dag
\end{bmatrix} = \frac{1}{\sqrt{2}}
\begin{bmatrix}
1 & +1 \\
1 & -1
\end{bmatrix}
\begin{bmatrix}
f_\Up \\ f_\Down^\dag
\end{bmatrix} ,
\end{equation}
the Hamiltonian~\eqref{eq:H} is rewritten as
\begin{multline}
\label{paper::eq:8}
H = \epsilon_f \sum_{\sigma=\Up,\Down} f_\sigma^\dag f_\sigma
+ U f_\Up^\dag f_\Up f_\Down^\dag f_\Down
+ \sum_k\epsilon_{k}c_{k\up}^\dag c_{k\up} \\{}
+ \sqrt\frac{\Gamma_F}{2\pi\rho_F}\sum_k
\left[c_{k\up}^\dag\left(f_\Up+f_\Down^\dag\right) + h.c.\right]
\end{multline}
with $\epsilon_f = \Delta_d-U/2$.  The Hamiltonian in \eqnref{paper::eq:8}
describes a single-orbital Anderson-type impurity level $\epsilon_f$ with
onsite interaction $U$, coupled to a spin-polarized conduction band with
strength $\Gamma_F/2$.
Despite the formal similarity, there are two important distinction between the model~\eqref{paper::eq:8} and the conventional single-impurity Anderson model:
(i) The model~\eqref{paper::eq:8} involves the pair tunneling, $c_{k\up}^\dag f_\Down^\dag$, which will turn out to play a crucial role below.
(ii) The spin index $\sigma=\Up,\Down$ for $f_\sigma$ indicates the spin direction along the spin $x$-direction whereas $\mu=\up,\down$ for $d_\mu$ along the spin $z$-direction.

% The system described by \eqnref{eq:H} can be more easily investigated under
% proper transformations. Here we consider two Bogoliubov transformations. The
% first one is to diagonalize the quantum dot Hamiltonian,
% \eqnref{eq:HSQD}. Under the Bogoliubov transformation
% \begin{equation}
% \label{eq:bt}
% \renewcommand\arraystretch{1.3}
% \begin{bmatrix}
% f_\Up \\ f_\Down^\dag
% \end{bmatrix}
% =
% \begin{bmatrix}
% \phantom{+}\cos\frac{\varphi}{2} & \sin\frac{\varphi}{2} \\
% - \sin\frac{\varphi}{2} & \cos\frac{\varphi}{2}
% \end{bmatrix}
% \begin{bmatrix}
% d_\up \\ d_\down^\dag
% \end{bmatrix},
% \end{equation}
% the components $H_\mathrm{SQD}$ and $H_\mathrm{T}$ in~\eqref{paper::eq:2} of the Hamiltonian are transformed into
% \begin{subequations}
% \begin{align}
% H_\mathrm{SQD}
% & = \epsilon_f \sum_{\sigma=\pm} n_\sigma + U n_+ n_- \\
% H_\mathrm{T}
% & = \sum_k \left[t_{F+} f_\Up^\dag c_{k\up} + t_{F-} f_\Down c_{k\up} + (h.c.)\right]
% \end{align}
% \end{subequations}
% with
% \begin{equation}
% \epsilon_f = \sqrt{\delta^2 + \Delta_d^2} - U/2,
% \
% \tan\varphi = \frac{\Delta_d}{\delta},
% \
% t_{F+} = t_F \cos\frac{\varphi}{2},
% \ \text{and}\
% t_{F-} = -t_F \sin\frac{\varphi}{2}.
% \end{equation}
% In this representation, both the dot degrees of freedom $d_\pm$ are
% tunnel-coupled to the lead, and more importantly, the induced superconductivity
% in the dot gives rise to the anomalous tunneling term, $f_\Down c_{k\up}$ between the
% dot and the lead.
% \begin{color}{red}

% \subsection{Relation to Other Models}

On the other hand, the particle-hole transformation
\begin{align}
d_1 = d_\up,
\quad
d_2 = d_\down^\dag,
\end{align}
transforms the model~\eqref{eq:H} to
\begin{multline}
\label{eq:HSQD:TLM}
H = - U (n_1 - 1/2) (n_2 - 1/2)
+ \Delta_d (d_1^\dag d_2 + d_2^\dag d_1) \\{}
+ \sum_k\epsilon_k c_{k\up}^\dag c_{k\down}
+ \sqrt{\frac{\Gamma_F}{\pi\rho_F}}\sum_k
(d_1^\dag c_{k\up} + c_{k\up}^\dag d_1) \,.
\end{multline}
In this model, the ferromagnetic lead is coupled to $d_1$ via a normal
tunneling and the pairing term has been transformed to a tunneling term between
dot orbital levels.  It is known as the resonant two-level system with
attractive interaction ($-U<0$) \cite{Zitko2009feb,Zitko2011nov}.

\subsection{Methods and Physical Quantities}
\label{paper::sec:2.2}

For a non-perturbative study of the many-body effects, we adopt the well-established
numerical renormalization group (NRG) method, which provides not only
qualitatively but also quantitatively accurate results for quantum impurity systems.
Specifically, we exploit the NRG method to identify the different phases of the system as well as to investigate their quantum transport properties.
Technically, we impose additional improvements, the
generalized Logarithmic discretization \cite{Campo2005sep,Zitko2009feb} with
the discretization parameter $\Lambda=2$ and the $z$-averaging
\cite{Yoshida1990may} with $N_z = 32$, on the otherwise standard
NRG procedure \cite{Wilson1975oct,Krishna-murthy80a,Bulla2008apr}.
We use the conduction band half-width $D=1$ as the unit of energy.

To identify the phases, we follow the (non-perturbative) renormalization group idea \cite{Wilson75a,Krishna-murthy80a,Krishna-murthy80b} and examine the conserved quantity
\begin{equation}
N_S = n_\up - n_\down + \sum_k c_{k\up}^\dag c_{k\up} - N_0
\end{equation}
of the ground state, where $N_0$ is the total charge number of the unperturbed
spin-polarized lead at zero temperature. Physically, $N_S$ is the
\emph{excess} spin number in the whole system.

The quantum transport properties of different phases and crossover regions are investigated by calculating the local spectral density and the charge relaxation resistance with the NRG method. The local spectral density (or local tunneling density of states) of the QD,
\begin{equation}
A_\mu(\omega)
= -\frac{1}{\pi\hbar} \im[G_\mu^R(\omega)] \,,
\end{equation}
is related to the Fourier transform $G_\mu^R(\omega)$ of the retarded Green's
function $G_\mu(t)$ for spin $\mu$,
$G_\mu(t) = -i\hbar\Theta(t) \Braket{\{d_\mu(t),d_\mu^\dag(0)\}}$.
The charge relaxation resistance $R_q(\omega)$ describes the response of the displacement current $I(t)$ through the QD in the presence of the ac gate voltage \cite{Buttiker1993jun,Buttiker1993sep,LeeMC2011may,LeeMC2014aug}. More explicitly, it is defined through the admittance $g(t) = (ie/\hbar) \Theta(t) \Braket{[I(t),n_d(t)]}$ by the relation
$1/g(\omega) = R_q(\omega) + i/\omega C_q(\omega)$, where $C_q(\omega)$ is the
quantum correction to the capacitance. The admittance in turn can be extracted from its
relation, $g(\omega) = i\omega (e^2/\hbar) \chi_c(\omega)$ to the dot charge
susceptibility $\chi_c(t) = -i\Theta(t)\Braket{[n_d(t),n_d]}$, which is
directly calculated with the NRG method.

\section{Results}
\label{paper::sec:3}

\Figref{fig:pd} shows the phase diagram which exhibits a quantum
phase transition between two phases, the spin singlet (S) and doublet (D)
phases, identified by the quantum number $N_S$ of the ground state calculated with the NRG method. Across the phase boundary, the quantum number $N_S$ of the ground state changes from $N_S=\pm1$ (doublet) to $N_S=0$ (singlet).
In addition, apart from the phase transition, we have found two
crossovers further distinguishing three regimes inside the singlet phase:
superconductivity-dominant ($\mathrm{S_S}$), mixed-valence (S$_\mathrm{M}$), and Kondo (S$_\mathrm{K}$) singlet regimes.
Below, we detail some interesting characteristics of each phase.

\subsection{Double Phase}
\label{paper::sec:3.1}

The doublet phase occupies the region of smaller $\Delta_d$ and $\Gamma_F$ of the phase diagram in Fig.~\ref{fig:pd}. The phase boundary is roughly linear for $\Gamma_F/U\ll 1/2$ as described by the equation
\begin{equation}
\label{paper::eq:6}
\Delta_d/U+\Gamma_F/U \approx 1/2 \,.
\end{equation}
Note that the ground state remains doubly degenerate with the excess spin
number $N_S=\pm1$ even in the presence of the coupling to the spin-polarized
ferromagnetic lead. It is due to the particle-hole symmetry. With the
particle-hole symmetry is broken, the degeneracy is lifted at finite
$\Gamma_F$ and the phase boundary is shifted accordingly \cite{endnote:1}.

\begin{figure}
\centering
\includegraphics[width=0.8\columnwidth]{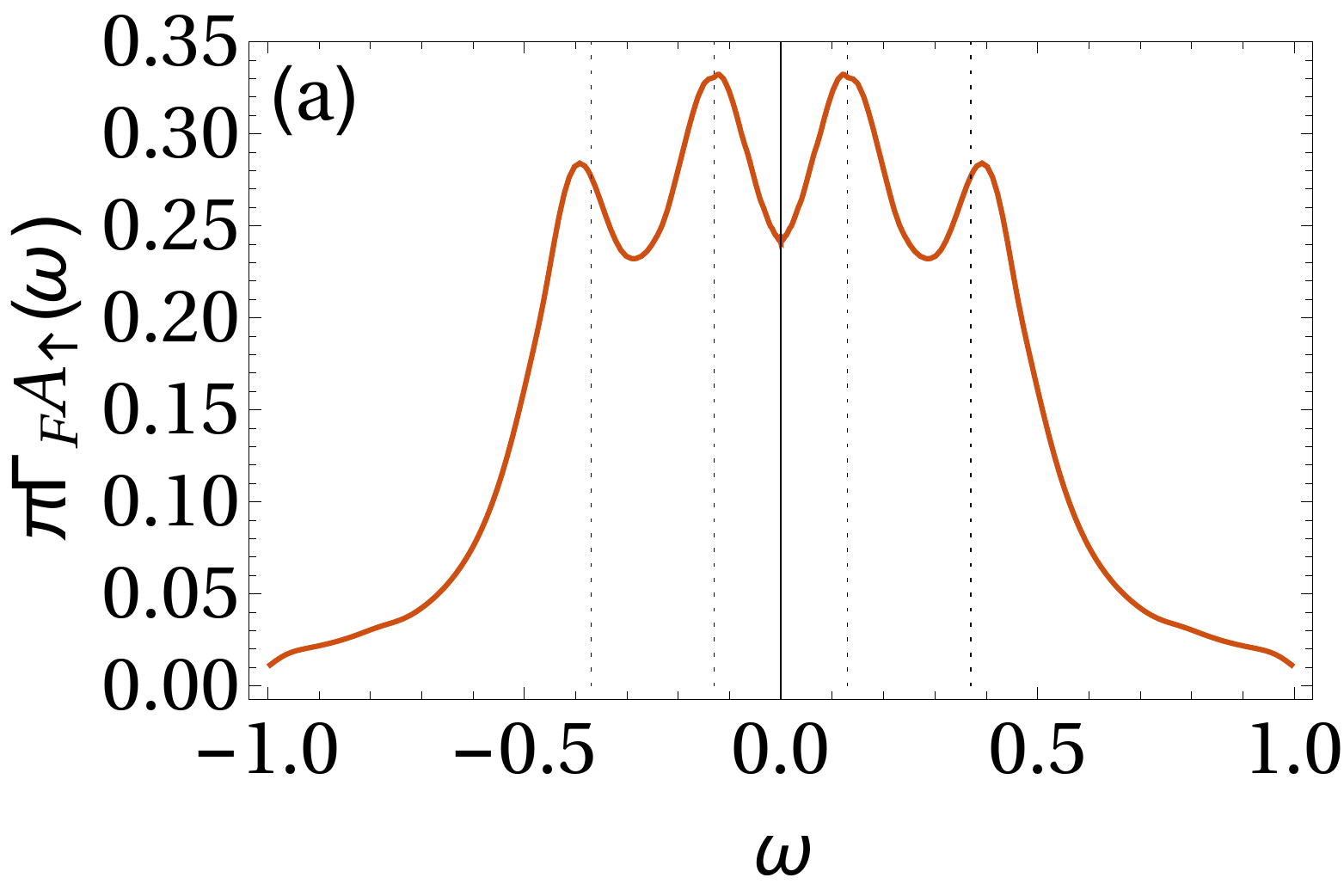} \\
\includegraphics[width=0.8\columnwidth]{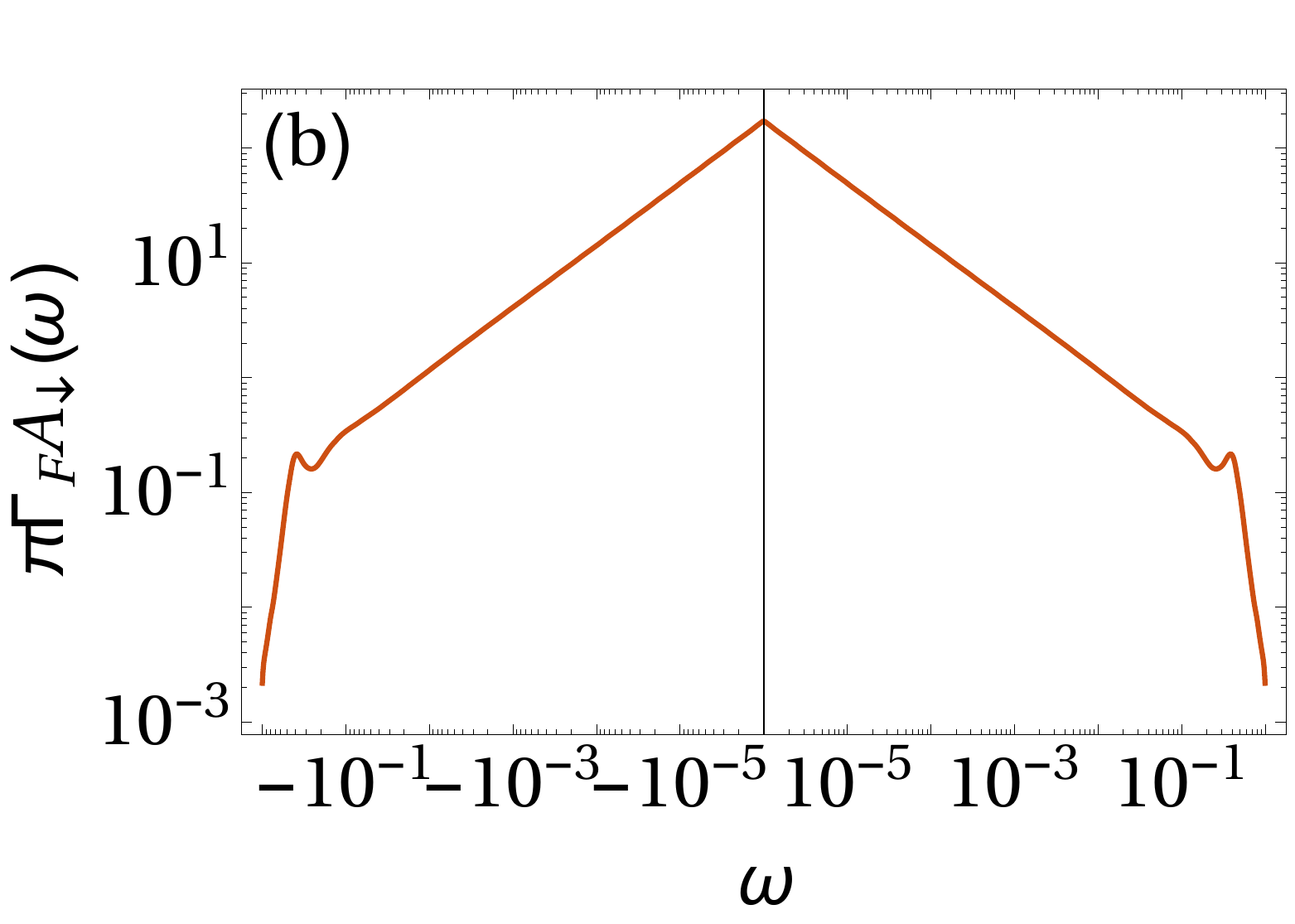} \\
\includegraphics[width=0.8\columnwidth]{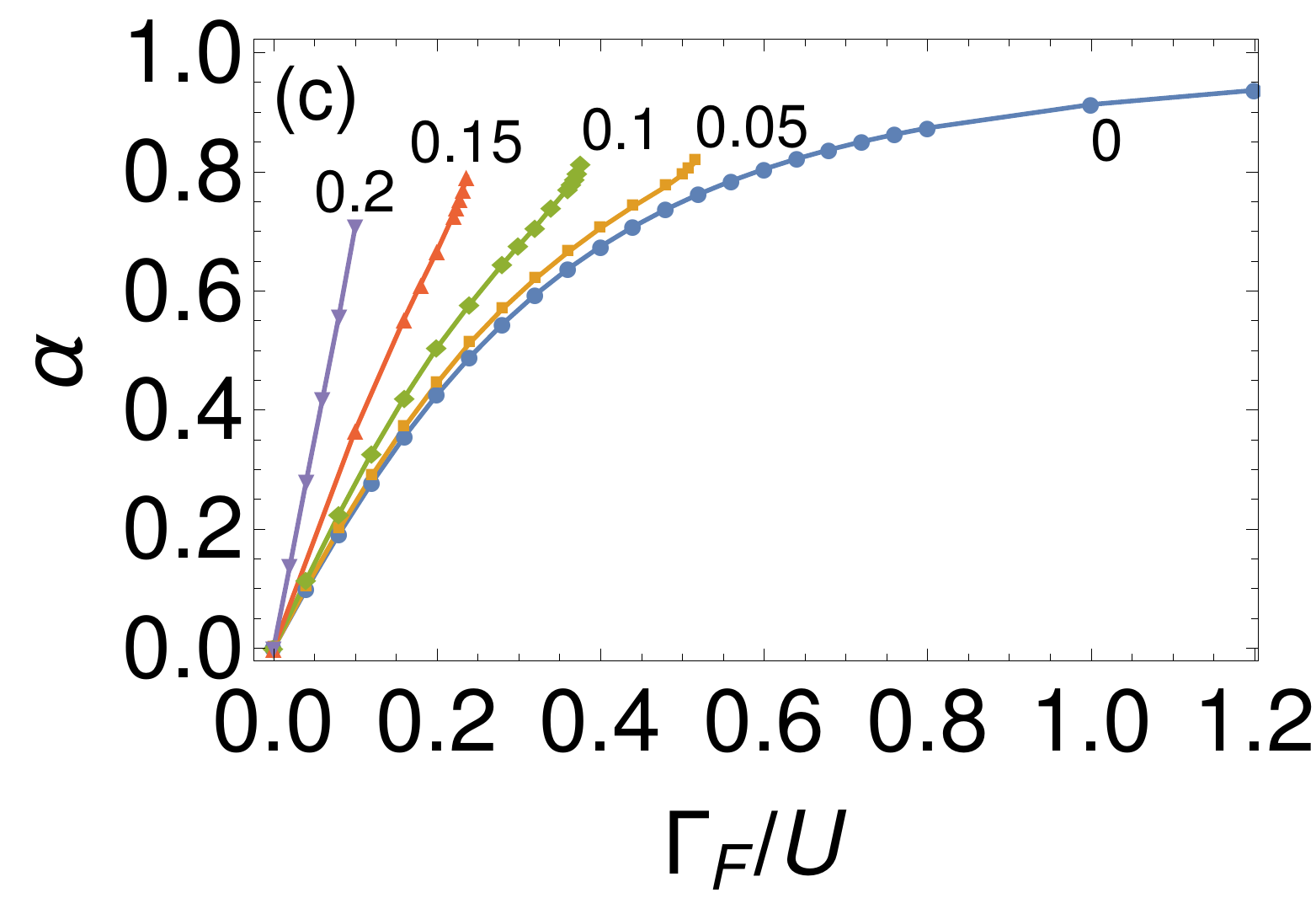}
\caption{(a,b) Spin-dependent spectral densities $A_\mu(\omega)$ in the spin doublet phase, corresponding to the point $1$ in \figref{fig:pd}. Here we have used $U=0.5D$,
  $\Gamma_F=0.1D$, and $\Delta_d=0.12D$. The dotted lines in (a) indicate the
  frequencies $|\hbar\omega| = U/2\pm\Delta_d$.  (c) The exponent $\alpha$ from
  the power-law relation of $A_\down(\omega)$. The line is a fitting
    curve for $\Delta_d=0$; see the text for the expression for it. The value
  of $\Delta_d/D$ are annotated.}
\label{fig:doublet}
\end{figure}

In the doublet phase, the local spectral densities $A_\mu(\omega)$ on the QD exhibit typical charge-fluctuation peaks at
\begin{math}
|\hbar\omega| \sim E_{S\pm}^0 - E_D^0 = U/2\pm\Delta_d;
\end{math}
see \figsref{fig:doublet}~(a) and (b).
Apart from those charge-fluctuation peaks, $A_\down(\omega)$ has an additional
power-law peak at the zero frequency $\omega = 0$, $A_\down(\omega) \propto |\omega|^{-\alpha}$ [see
\figref{fig:doublet}~(b)].  This power-law peak at the zero energy suggests
that the doublet phase is `marginal' in the RG sense.
The exponent $\alpha$ is found to increase monotonically with increasing $\Gamma_F$
and $\Delta_d$, and is well fitted to $\alpha = 1 - (2/\pi)\tan^{-1}(U/2\Gamma_F)$ for small $\Delta_d$ [see \figref{fig:doublet}~(c)].

\subsection{Singlet Phase: Superconductivity-Dominant Singlet}
\label{paper::sec:3.2}

For larger values of $\Delta_d$,\footnote{Recall that the proximity-induced pairing potential $\Delta_d\sim\Gamma_S$. Therefore, the large-$\Delta_d$ limit corresponds to the strong coupling to the superconductor in the original system in Fig.~\ref{fig:pd} (a).} the system has a singlet ground state. In particular, the region of larger $\Delta_d/U$ and smaller $\Gamma_F/U$ of the phase diagram Fig.~\ref{fig:pd} is characterized by the strong Cooper pairing. It is natural as the ground state of the unperturbed QD ($\Gamma_F=0$) is the spin singlet $\ket{S_-^0}$ composed of empty or doubly occupied states [see \eqnref{eq:iQD}] due to the proximity-induced superconductivity.
Such superconductivity-dominant singlet region is separated from other singlet regions by a crossover boundary, roughly described by the equation [cf.~\eqnref{paper::eq:6}]
\begin{equation}
\label{paper::eq:7}
\Delta_d/U - \Gamma_F/U \approx 1/2 \,.
\end{equation}

\begin{figure}
\centering
\includegraphics[width=0.8\columnwidth]{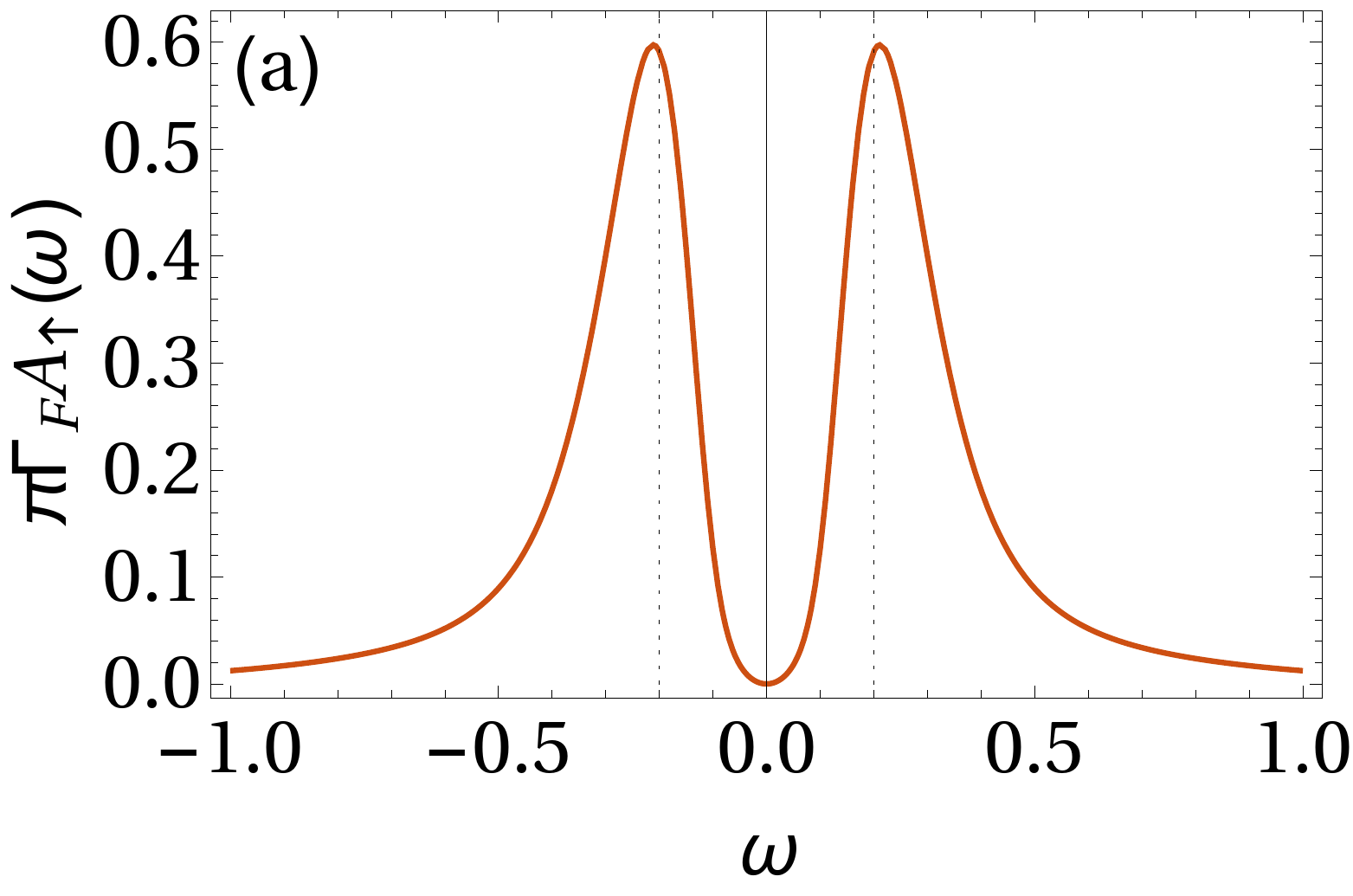}
\includegraphics[width=0.8\columnwidth]{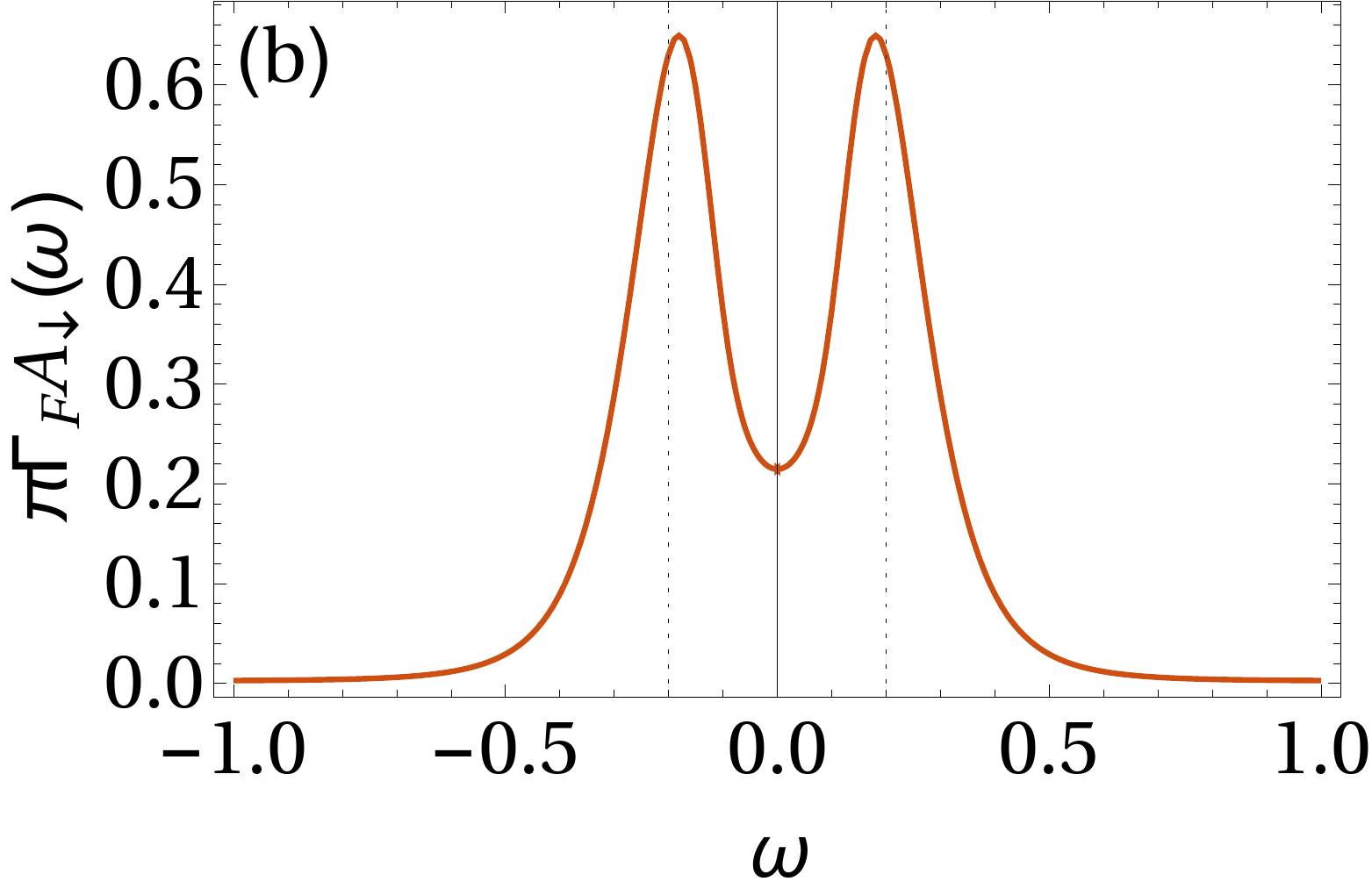}
\caption{Dot spectral densities in the superconductivity-dominant singlet
  regime corresponding to the point $3$ in \figref{fig:pd}. Here we have used
  $U=0.5D$, $\Gamma_F=0.4D$, and $\Delta_d=0.45D$. The dotted lines indicate
  the frequencies $|\hbar\omega| = \Delta_d-U/2$. In (a), the spectral density vanishes at zero frequency due a Fano-like destructive interference.}
\label{fig:SuperSinglet}
\end{figure}

Because in this regime the superconductivity prevails over all the other types of correlations, the dot spectral densities [see Figs.~\ref{fig:SuperSinglet}~(a) and (b)] are simply given by the charge fluctuation peaks at $|\hbar\omega| \sim E_D^0 - E_{S-}^0 = \Delta_d - U/2$, broadened by the weak tunnel coupling $\Gamma_F$.

However, there is one noticeable feature in the spin-up spectral density $A_\up(\omega)$. That is, $A_\up(\omega=0)=0$ exactly, which is the consequence of the Fano-like destructive interference between two kinds of
dot-lead tunneling processes. It will be discussed in detail in Section~\ref{paper::sec:4.3}.

\subsection{Singlet Phase: Mixed-Valence Singlet}
\label{paper::sec:3.3}

The most interesting singlet phase occurs near $\Delta_d/U\approx 1/2$ with
finite $\Gamma_F/U$ in the phase diagram (Fig.~\ref{fig:pd}). We call it a ``mixed-valence singlet'' region because $\epsilon_f<\Gamma_f$ in the model~\eqref{paper::eq:8} regarding $\epsilon_f$ and $U$ as independent parameters; see the further discussions in Section~\ref{paper::sec:4.4}. It is distinguished from the doublet phase by the true phase boundary~\eqref{paper::eq:6} and separated from the superconductivity-dominant singlet state by the crossover boundary~\eqref{paper::eq:7}; that is,
\begin{equation}
|\Delta_d/U-1/2|\approx\Gamma_F/U .
\end{equation}
It is also separated from still another singlet state for $\Gamma_F/U\gg 1$,
which is characterized by the Kondo behaviors (see also
Section~\ref{paper::sec:3.4}), by another crossover.

\begin{figure}
\centering
\includegraphics[width=0.8\columnwidth]{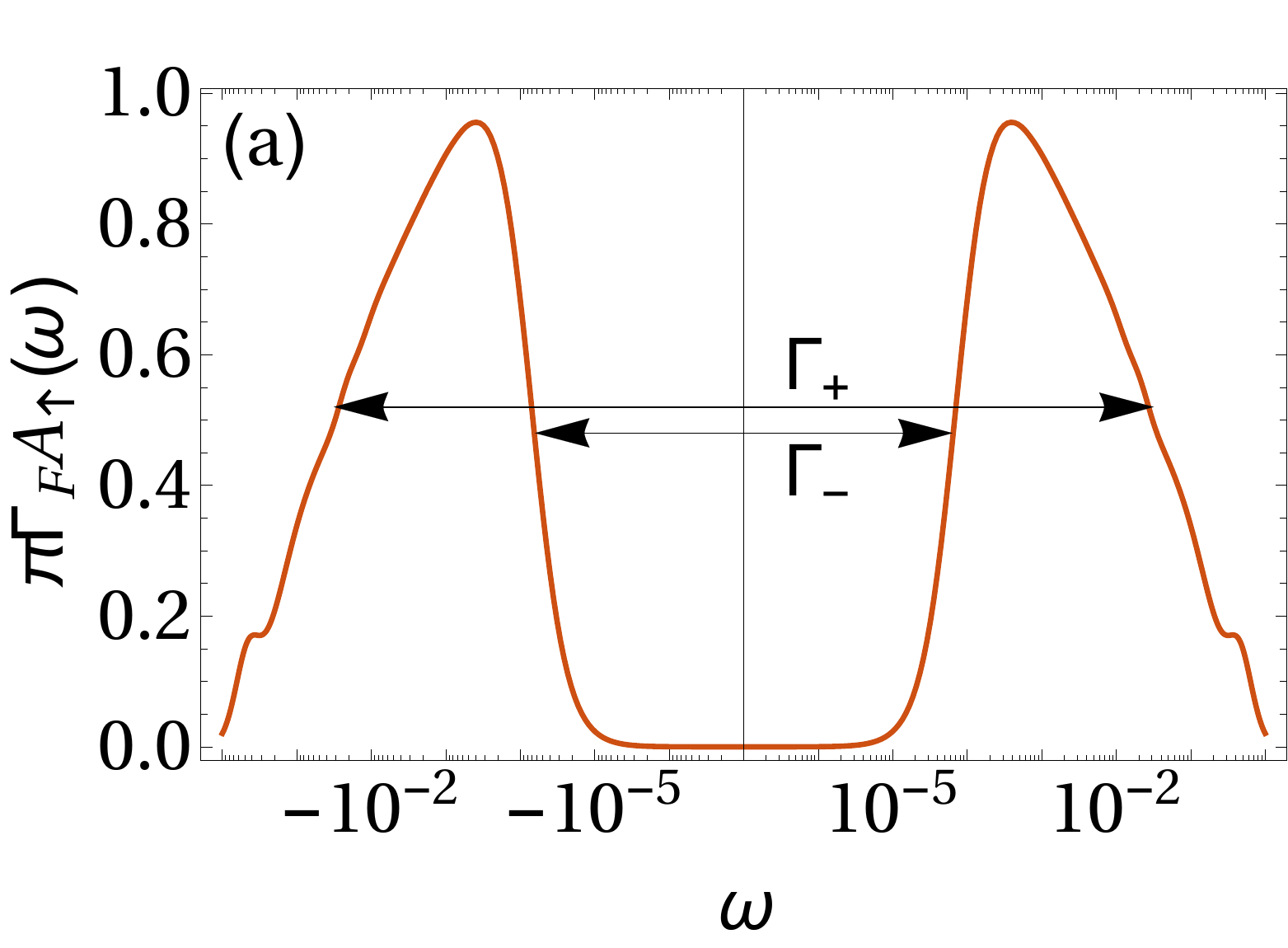}
\includegraphics[width=0.8\columnwidth]{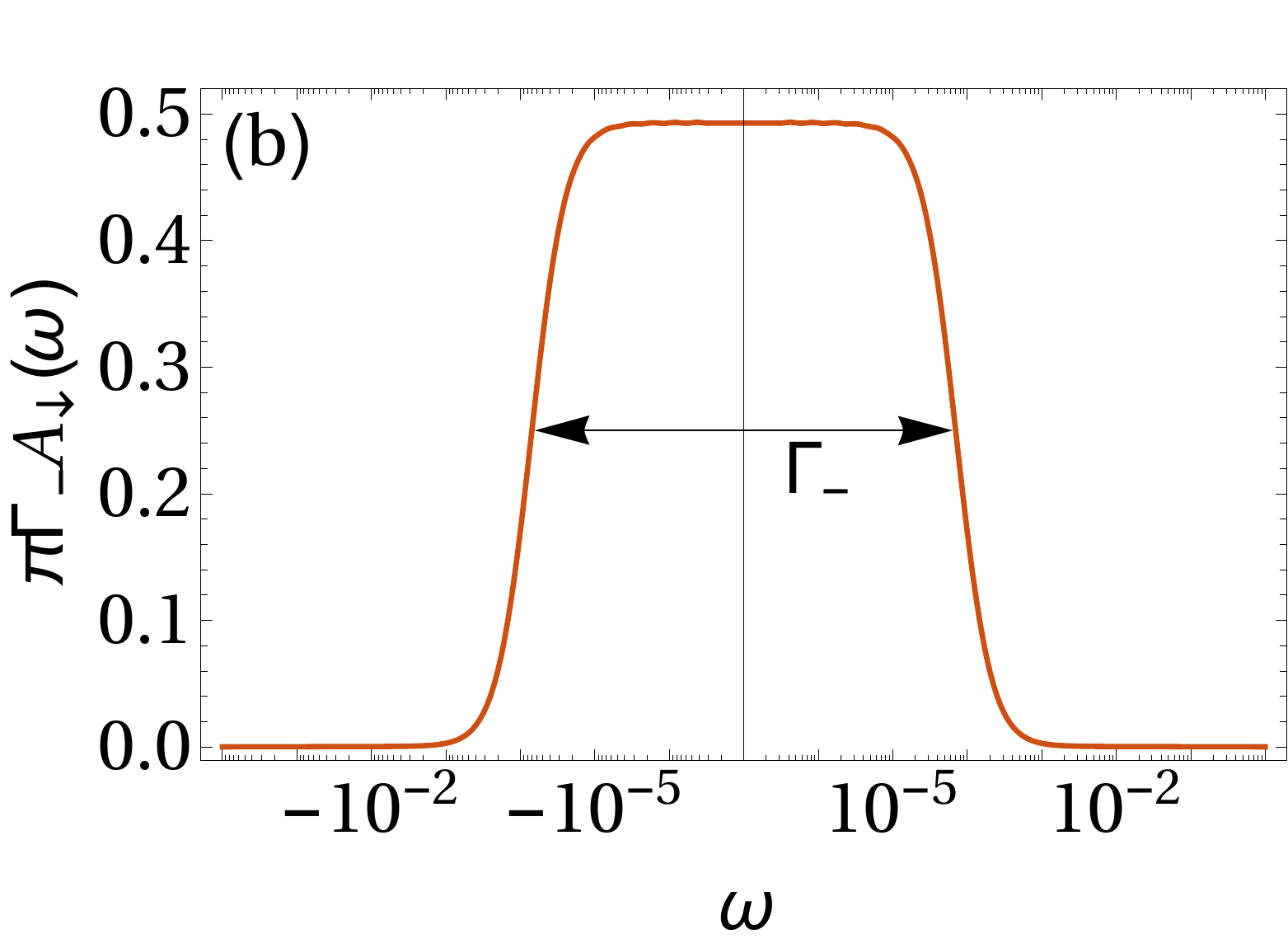}
\caption{Dot spectral densities in the mixed-valence singlet regime at
  the point $4$ in \figref{fig:pd}. Here we have used $U=0.5D$,
  $\Gamma_F=0.4D$, and $\Delta_d=0.19D$.}
\label{fig:mixedvalence}
\end{figure}

The two spin-dependent spectral densities $A_\mu(\omega)$ in the mixed-valence
singlet state, as shown in Fig.~\ref{fig:mixedvalence}, put stark contrast with each other: While $A_\down(\omega)$ for
the minority spin features a usual Lorentzian peak of width $\Gamma_-$ at the
zero frequency, $A_\up(\omega)$ for the majority spin has a Lorentzian dip of
the same width $\Gamma_-$ superimposed on a broader peak structure of width
$\Gamma_+$. Later [see Section~\ref{paper::sec:4.4}], we will attribute this
dip structure to a destructive interference between two different types of
tunneling processes based on an effective non-interacting theory.

\subsection{Singlet Phase: Kondo Singlet}
\label{paper::sec:3.4}

When the QD couples strongly with the spin-polarized lead ($\Gamma_F/U\gg1,\Delta_d/U$), the system displays still another type of singlet correlation. We call this state as a Kondo singlet state as it corresponds to the so-called `charge Kondo state' \cite{Matveev91a,Iftikhar15a}; see Section~\ref{paper::sec:4.5}. In the charge Kondo state, the excess charge on the QD plays the role of a pseudo-spin.

\begin{figure}
\centering
\includegraphics[width=0.7\columnwidth]{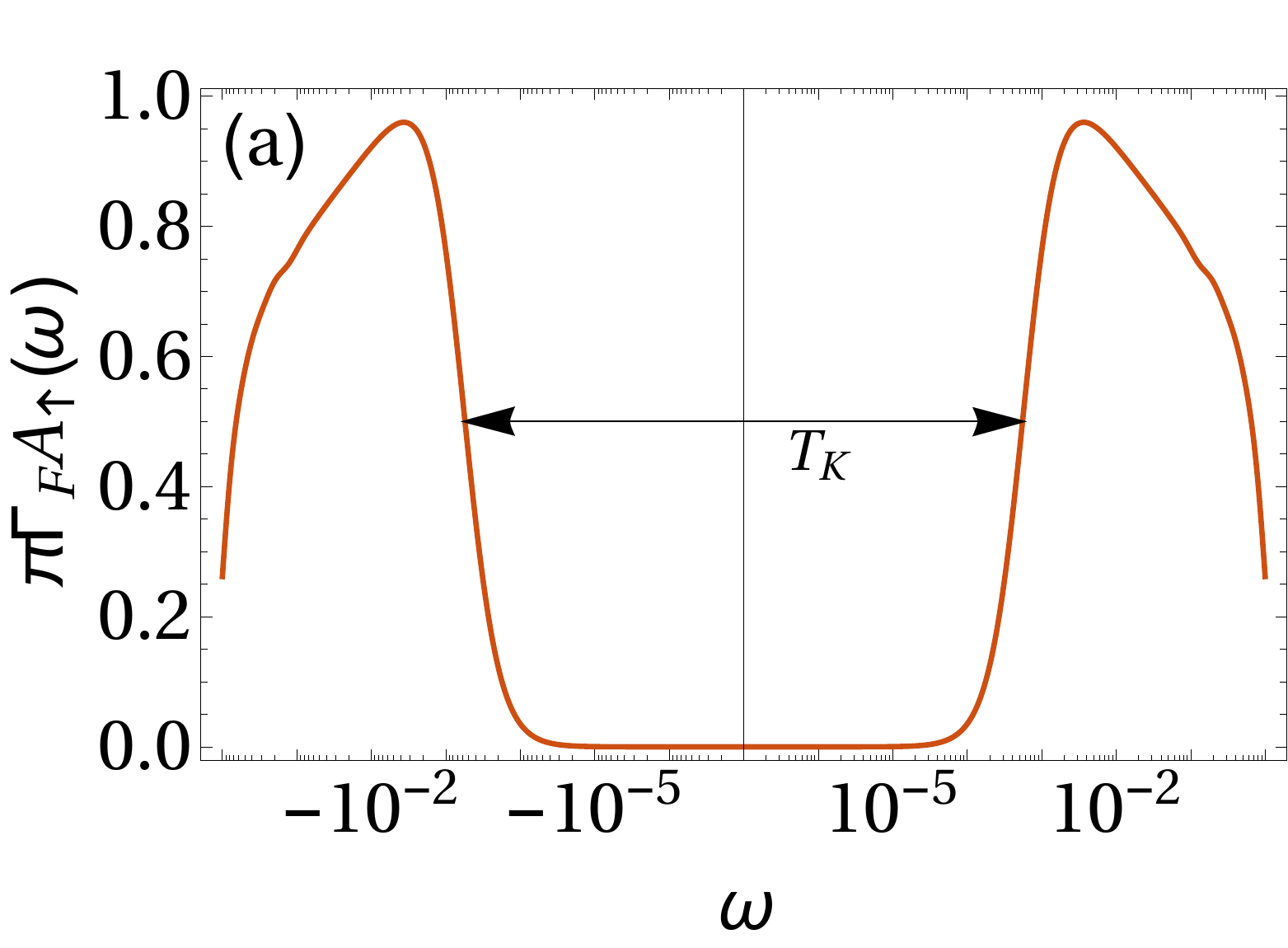}
\includegraphics[width=0.7\columnwidth]{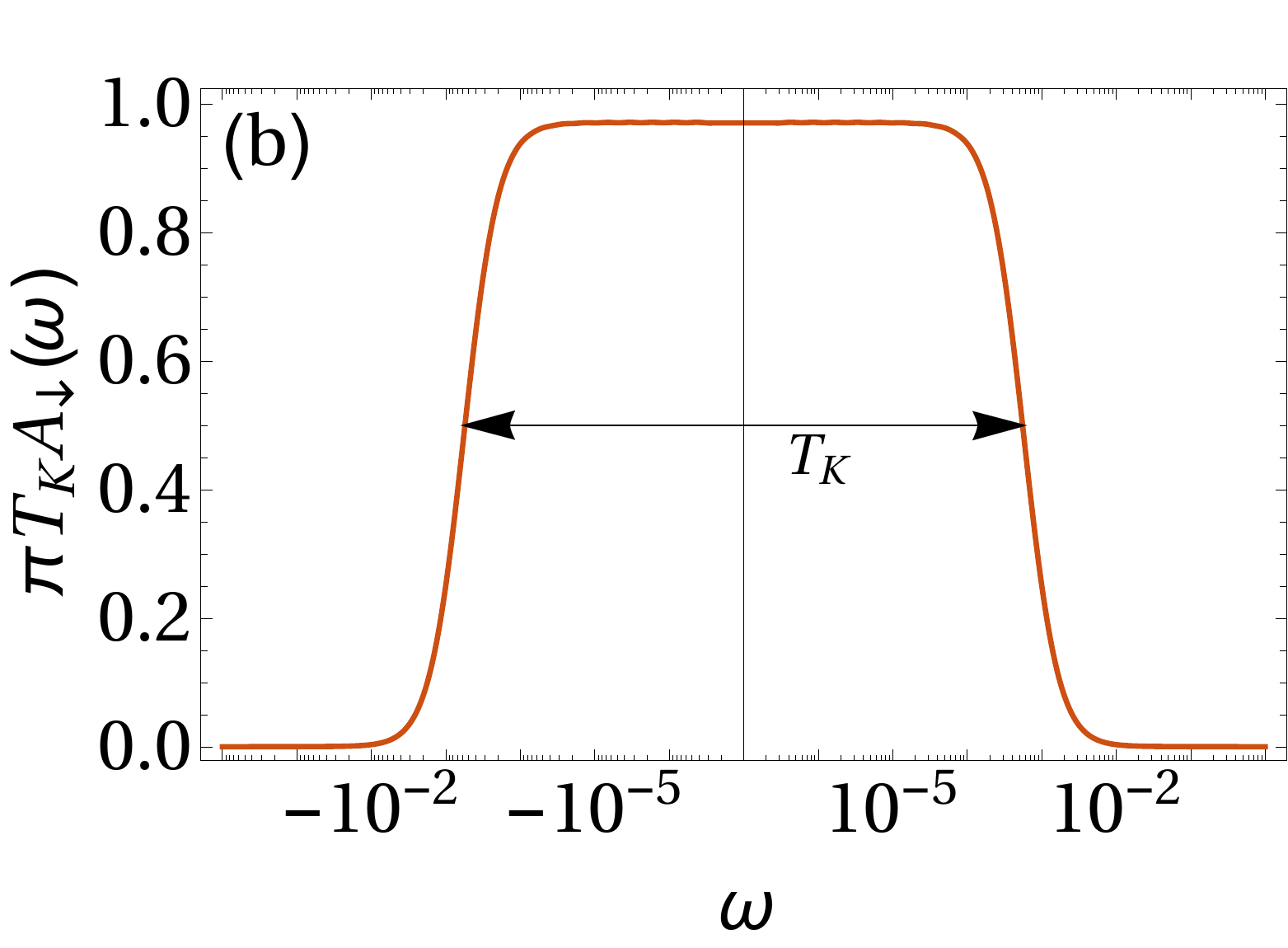}
\includegraphics[width=0.7\columnwidth]{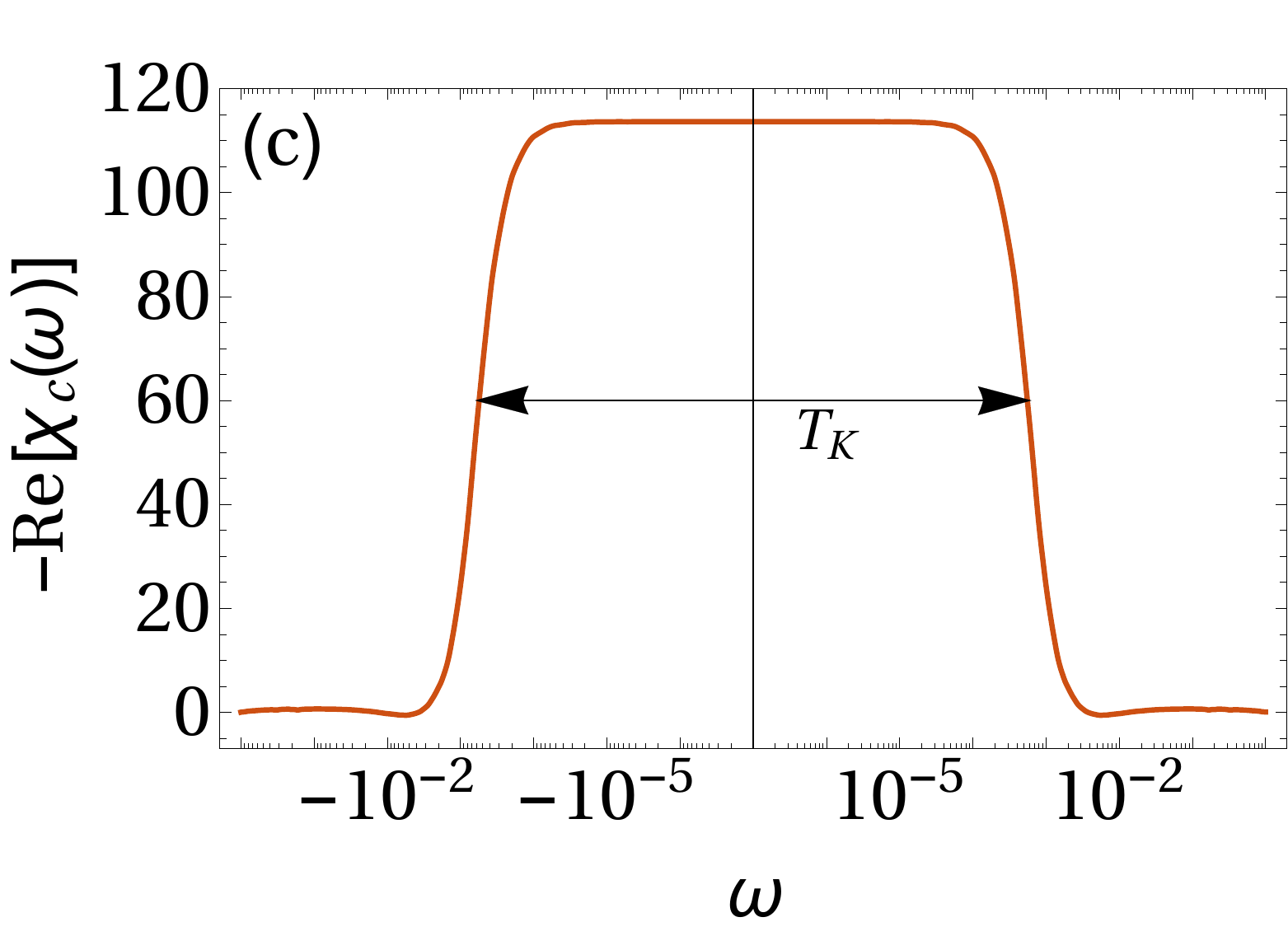}
\includegraphics[width=0.7\columnwidth]{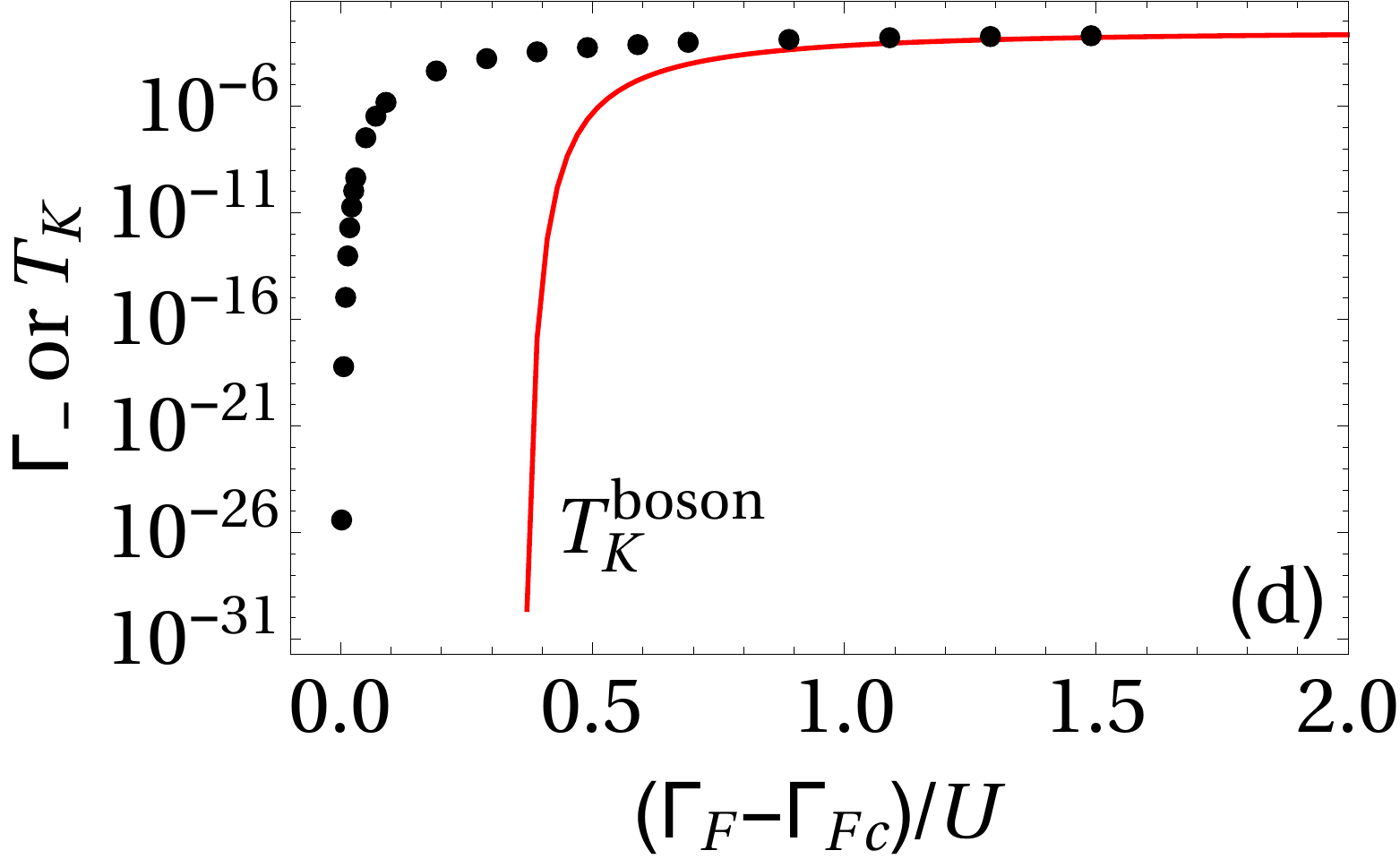}
\caption{(a,b,c) Dot spectral densities and charge susceptibility in the Kondo
  regime of the singlet phase at point $2$ in \figref{fig:pd}. Here we have
  used $U=0.5D$, $\Gamma_F=0.4D$, and $\Delta_d=0.125D$. (d) The width of the
  central peak of $A_\down(\omega)$ and $T_K^\mathrm{boson}$ from
  \eqnref{eq:tk:boson} at $\Delta_d=0.125D$.}
\label{fig:Kondo}
\end{figure}

As shown in Fig.~\ref{fig:Kondo}, the peak shapes of the spectral densities $A_\mu(\omega)$ are similar to those
in the mixed-valence singlet state described in Section~\ref{paper::sec:3.3}. The dip
structure in $A_\up(\omega)$ for the majority spin is again attributed to the
Fano-like destructive interference. However, the normalized peak height
$\pi T_KA_\down(\omega)$ for the minority spin is now unity, demonstrating the
charge Kondo effect; {the peak height of $\pi\Gamma_-A_\down(\omega=0)$
  grows from zero to unity as one moves from the mixed-valence regime to the
  Kondo regime} [compare \figref{fig:Kondo} (b) with \figref{fig:mixedvalence}
(b)].  Further, the peak width of $A_\down(\omega)$, or the dip width of
$A_\up(\omega)$, is identified as the charge Kondo temperature $T_K$.

The charge Kondo effect is also manifested in the charge susceptibility $\chi_c(\omega)$ of the QD shown in \figref{fig:Kondo}~(c). Its real part displays a pronounced central peak of the same width $T_K$. In the conventional (spin) Kondo effect, this susceptibility corresponds to the spin susceptibility.

\section{Discussion}
\label{sec:discussion}

The NRG calculations reported in the previous section clearly display a quantum phase transition between the spin singlet and doublet phases. Here we use some analytical but approximate methods to understand deeper the nature of the transition and the characteristics of the different phases.

As seen in the equivalent model~\eqref{paper::eq:8}, our system is described by
a generalized form of the Anderson impurity model. The Anderson impurity model
\cite{Anderson61a} has been studied in various theoretical methods; using the
variational method \cite{Varma76a}, the scaling theory
\cite{Jefferson77a,Haldane78a}, the numerical renormalization group method
\cite{Krishna-murthy80b}, and the $1/N$ expansion \cite{Ramakrishnan82a}.
Here we extend some of these methods.

\bigskip

\subsection{Mixed-Valence Transition}
\label{paper::sec:4.1}

We first examine analytically the phase boundary between the doublet and
singlet phases found in Section~\ref{paper::sec:3} based on the NRG method. Our
analysis consists of two steps depending on the relevant energy scale. At
higher energies (the band cutoff $\Lambda\gtrsim\Gamma_F$),\footnote{The band cutoff $\Lambda$ here is not to be confused with the band discretization parameter of the NRG in Section~\ref{paper::sec:2.2}.} we extend the scaling theory
\cite{Jefferson77a,Haldane78a} to integrate out the high-energy excitations. At
lower energies ($\Lambda<\Gamma_F$), we extend the variational method \cite{Varma76a}.

Following Haldane's scaling argument \cite{Jefferson77a,Haldane78a}, it is
straightforward to integrate out the high energy states in the conduction band up
to $\Gamma_F$ and keep track of the scaling of the parameters $\epsilon_f$ and
$U$ in the equivalent model~\eqref{paper::eq:8}; concerning the model~\eqref{paper::eq:8} it is convenient to regard $\epsilon_f$ and $U$ (rather than $\Delta_d$ and $U$) as independent parameters.  We found that even though our
system has only a single spin channel the anomalous tunneling term acts as the
tunneling via the second spin channel so that the scaling result is exactly the
same as the one for the conventional Anderson model:
\begin{equation}
\epsilon_f(\Lambda)
= \epsilon_f^* - \frac{\Gamma_F}{\pi} \ln\frac{\Lambda}{\Gamma_F}
\end{equation}
with the scaling invariant
$\epsilon_f^*=\epsilon_f(\Lambda=\Gamma_F)$ and the band cutoff
$\Lambda$. Therefore, as in the conventional Anderson impurity model, it is possible to identify three regimes:
the empty/doubly-occupied ($|\epsilon_f^*|\gg\Gamma_F$), the
mixed-valence ($|\epsilon_f^*|\lesssim\Gamma_F$), and the local-moment
regimes ($\epsilon_f^*\ll-\Gamma_F$).
For the conventional Anderson impurity model, in all these regimes
the renormalization beyond the Haldane's scaling eventually flows into the spin
singlet state, so there are only crossovers between the regimes.
However, for our system the local-moment
regime does not flow into the singlet state because there is only a single spin channel and the anomalous tunneling
term prevents the formation of the conventional Kondo correlation. Therefore, a
transition takes place between the mixed-valence and local-moment regimes;
hence the transition is named as the mixed-valence one.

To see this more clearly,\footnote{From the numerical point of view, the disappearance of the Kondo correlation in the local-moment regime is already well implemented by the non-perturbative NRG method.} we extend the variational method. Here we focus on the case of $U\to\infty$. This condition rules out the doubly occupied state on the QD (recall that concerning the model~\eqref{paper::eq:8} $\epsilon_f$ and $U$ are regarded as independent parameters) and makes the variational analysis much simpler; the finite $U$ should involve more states but would not alter the main qualitative feature of the transition found in the $U\to\infty$ case.
We take a variational ansatz for the ground states in spin singlet and doublet states, respectively, up to the second order in the dot-lead tunneling
\begin{widetext}
\begin{subequations}
\label{eq:ansatz}
\begin{align}
\ket{S}
& =
\left[
\alpha_0
+ \sum_{k<k_F} \alpha_{k+} f_\Up^\dag c_{k\up}
+ \sum_{k>k_F} \alpha_{k-} c_{k\up}^\dag f_\Down^\dag
+ \sum_{k>k_F,k'<k_F} \alpha_{kk'} c_{k\up}^\dag c_{k'\up}
\right] \ket{\mathrm{FS}}_0 \\
\ket{D_\up}
& =
\left[
\beta_0 f_\Up^\dag + \sum_{k>k_F} \beta_k c_{k\up}^\dag
+ \sum_{k>k_F,k'<k_F}
\left( \beta_{kk'+} f_\Up^\dag c_{k\up}^\dag c_{k'\up}
+ \beta_{kk'-} c_{k\up}^\dag c_{k'\up}^\dag f_\Down^\dag\right)
\right]
\ket{\mathrm{FS}}_0,
\end{align}
\end{subequations}
where $\ket{\mathrm{FS}}_0$ is unperturbed Fermi sea and $k_F$ is the Fermi wave
number.
The states satisfy the normalization condition,
$\Braket{S|S} = \Braket{D_\up|D_\up} = 1$. The coefficients $\alpha$ and $\beta$ in
these two states are to be determined by the minimization condition of the energy expectation value with respect to these states:
$\Braket{S|H|S} := E_0 + \epsilon_f + \epsilon_S$ and
$\Braket{D_\up|H|D_\up} := E_0 + \epsilon_f + \epsilon_D$, where $E_0$ is
the unperturbed energy of $\ket{\mathrm{FS}}_0$. By applying the Lagrange multiplier
method under the normalization constraint, we obtain the coupled differential
equations:
\begin{subequations}
\begin{align}
\epsilon_S \alpha_0
& =
- \epsilon_f \alpha_0
+ \frac{t_F}{\sqrt2} \left(\sum_{k<k_F} \alpha_{k+} - \sum_{k>k_F} \alpha_{k-}\right)
\\
\epsilon_S \alpha_{k+}
& =
\frac{t_F}{\sqrt2} \alpha_0 - \epsilon_k \alpha_{k+}
+ \frac{t_F}{\sqrt2} \sum_{k'>k_F} \alpha_{k'k}
\\
\epsilon_S \alpha_{k-}
& =
- \frac{t_F}{\sqrt2} \alpha_0 + \epsilon_k \alpha_{k-}
+ \frac{t_F}{\sqrt2} \sum_{k'<k_F} \alpha_{kk'}
\\
\epsilon_S \alpha_{kk'}
& =
\frac{t_F}{\sqrt2} \left(\alpha_{k'+} + \alpha_{k-}\right)
+ (\epsilon_k - \epsilon_{k'} - \epsilon_f) \alpha_{kk'}
\end{align}
\end{subequations}
and
\begin{subequations}
\begin{align}
\epsilon_D \beta_0
& = \frac{t_F}{\sqrt2} \sum_{k>k_F} \beta_k
\\
\epsilon_D \beta_k
& =
\frac{t_F}{\sqrt2} \beta_0 + (\epsilon_k - \epsilon_f) \beta_k
+
\frac{t_F}{\sqrt2}
\left[\sum_{k'<k} \beta_{k'k-} - \sum_{k'>k} \beta_{kk'-} - \sum_{k'<k_F} \beta_{kk'+}\right]
\\
\epsilon_D \beta_{kk'+}
& = - \frac{t_F}{\sqrt2} \beta_k + (\epsilon_k - \epsilon_{k'}) \beta_{kk'+}
\\
\epsilon_D \beta_{kk'-}
& =
- \frac{t_F}{\sqrt2} (\beta_k - \beta_{k'}) + (\epsilon_k + \epsilon_{k'}) \beta_{kk'-}.
\end{align}
\end{subequations}
\end{widetext}

\begin{figure}
\centering
\includegraphics[width=0.8\columnwidth]{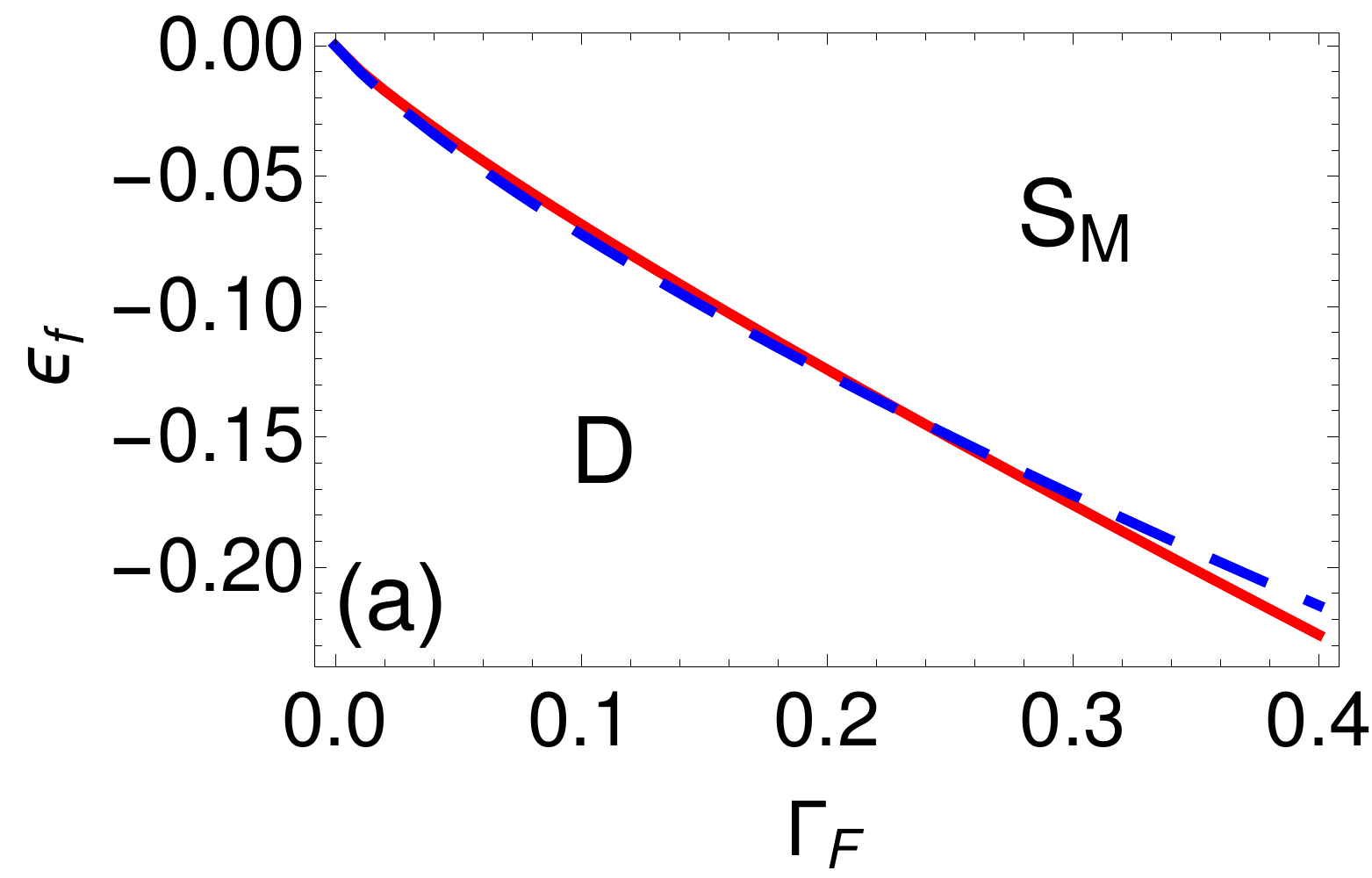}\quad%
\includegraphics[width=0.8\columnwidth]{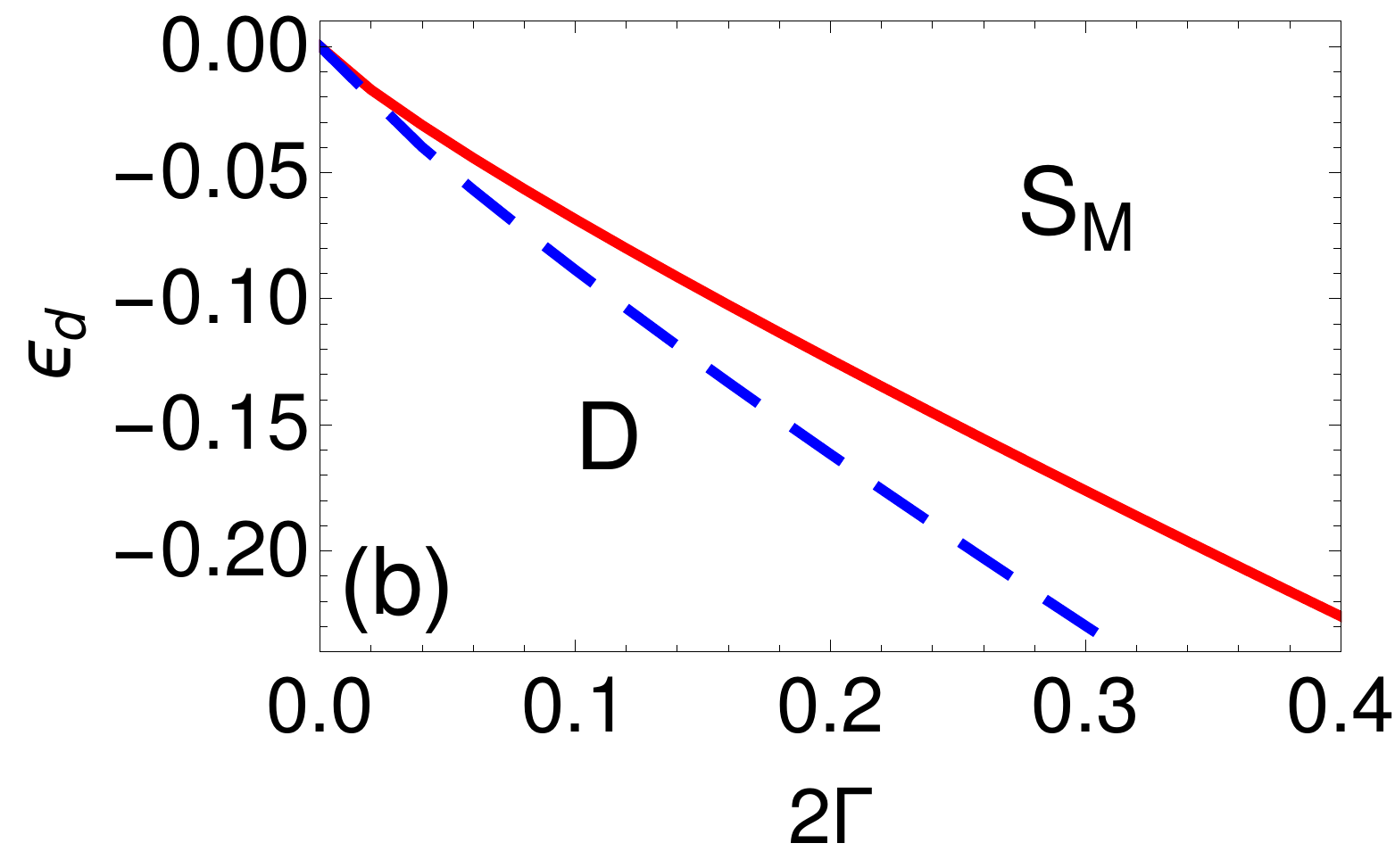}
\caption{Phase diagrams obtained from the variational method in the
  $U\to\infty$ limit (a) for our model and (b) for conventional
  single-impurity Anderson model. The solid and dotted lines are phase
  boundaries when up to the first and second-order terms are taken into
  account, respectively.}
\label{fig:vm}
\end{figure}

Up to the first order (by setting $\alpha_{kk'} = \beta_{kk'\pm} = 0$), the
equations for $\epsilon_S$ and $\epsilon_D$ can be obtained in closed form:
\begin{subequations}
\label{eq:vm1}
\begin{align}
\epsilon_S
& =
- \epsilon_f
- \frac{\Gamma_F}{\pi} \ln\left(1 + \frac{D}{|\epsilon_S|}\right)
\\
\epsilon_D
& =
-
\frac{\Gamma_F}{2\pi}
\ln\left(1 + \frac{D}{|\epsilon_D| - \epsilon_f}\right).
\end{align}
\end{subequations}
These equations can be solved numerically, and two different phases, in each of
which either $\epsilon_S < \epsilon_D$ or $\epsilon_S > \epsilon_D$, are
identified, as shown in \figref{fig:vm}~(a). Although a closed form equations
for $\epsilon_S$ and $\epsilon_D$ are not available with the second-order terms
included, the whole differential equation can be solved numerically by
discretizing the lead dispersion. It is found that the inclusion of the
second-order terms hardly changes the phase boundary. On similar reasoning, one can see that the phase boundary remains intact upon including the higher-order terms in the variational wave functions.

It is in stark contrast with the similar variational analysis for the conventional
Anderson impurity model in Appendix~\ref{sec:vmsiam}: Up to the
first-order the equations for $\epsilon_S$ and $\epsilon_D$ are the same as those
for our models [see \eqnref{eq:vm:siam}]. Therefore, at this order a phase transition
between the spin singlet and doublet states also takes place even in the
conventional Anderson impurity model. This apparent contradiction to the
well-known fact that the ground state of the conventional Anderson impurity
model is always spin singlet is due to the perturbative construction of the
ansatz.
As illustrated in \figref{fig:vm}~(b),
the spin doublet region shrinks for the
conventional Anderson model when one includes the higher-order terms.
In other words, the Kondo ground state involves all the higher-order singlet states
between the dot and the lead~\cite{Gunnarsson83a}.

This difference can be inferred from the comparison between two ansatz, \eqnsref{eq:ansatz} and (\ref{eq:ansatz:siam}).
For the spin singlet state, the number of the
particle-hole excitations in the second-order term for our model is by half
smaller than that for the conventional Anderson impurity model because of the difference in the channel
numbers. On the other hand, it is not the case for the doublet state. It
explains why the singlet state in our model does not lower its energy upon
including the higher-order terms, compared to the doublet state, and also why
the Kondo correlation cannot arise.

\subsection{Doublet Phase}
\label{paper::sec:4.2}

We now investigate the characteristics of the different phases (and subregions inside the singlet phase). We start with the doublet phase by applying the Schrieffer-Wolff transformation on
the assumption that $\Gamma_F \ll |\epsilon_f|, U$.
The model~\eqref{paper::eq:8} is then
transformed to an effective Kondo-like model:
\begin{equation}
\label{eq:Heff}
H \approx
H_\mathrm{eff}
= J \bfs\cdot\bfS + \sum_k \epsilon_k c_{k\up}^\dag c_{k\up} \,.
\end{equation}
Here the impurity spin-1/2 operator $\bfS$ is defined by
\begin{equation}
S^+ = (S^-)^\dag = \ket\Up \bra\Down \,,\quad
S^z = \ket\Up\bra\Up - \ket\Down\bra\Down \,,
\end{equation}
where
\begin{equation}
\ket\sigma = f_\sigma^\dag\ket{0} \quad (\sigma=\Up,\Down) \,.
\end{equation}
On the other hand, the conduction-band spin,
$\bfs = \sum_{kk'\nu\nu'} \psi_{k\nu}^\dag \bftau_{\nu\nu'} \psi_{k'\nu'}$ is
defined over the two-component Nambu spinor $\psi_{k\nu}$ with
$\psi_{k1} = c_{k\up}$ and $\psi_{k2} = c_{k\up}^\dag$ with $\bftau$ being the
Pauli matrices in the Nambu space (i.e. the particle-hole isospin space). The
isotropic exchange coupling is obtained as
$\rho_F J \approx \Gamma_F/\pi|\epsilon_f|$.  The model \eqref{eq:Heff} is
formally the same as the usual Kondo model except the fact that the conduction
spin is replaced by the isospin in the Nambu space. This replacement, however,
makes a crucial difference in poor man's scaling \cite{Anderson70a,Krishna-murthy80a,Krishna-murthy80b}. For example, the typical
scaling of $J_z$ term vanishes at least up to the second order:
\begin{equation}
- 2 \frac{J_\perp^2}{\epsilon_{k_3}}
\left(
c_{k_1} \dot{c}_{k_2} \dot{c}_{k_3}^\dag c_{k_4}^\dag
+
\dot{c}_{k_2} c_{k_1} \dot{c}_{k_3}^\dag c_{k_4}^\dag
\right) \ket\up \bra\up
\approx 0 \,.
\end{equation}
% (see \secref{sec:pm:doublet} for details).
These results imply that unlike the true Kondo model involving real spins, the
exchange coupling in \eqnref{eq:Heff} involving particle-hole isospins is
marginal in the RG sense. Namely, it does not scale as one goes down to lower
energies.  The NRG results discussed in Section~\ref{paper::sec:3.1} support
this scaling analysis.

\subsection{Singlet Phase: Superconductivity-Dominant Singlet}
\label{paper::sec:4.3}

The superconductivity-dominant singlet phase can be easily understood within the perturbative argument. When the QD is isolated ($\Gamma_F=0$), the pairing potential $\Delta_d$ dominates over the on-site interaction $U$ for $\Delta_d/U>1/2$; see Eq.~\eqref{eq:iQD}. As the tunneling coupling $\Gamma_F$ is turned on, the above feature does not change qualitatively unless $\Gamma_F$ exceeds $\Delta_d$ significantly. As $\Gamma_F/U$ grows further beyond $\Delta_d/U-1/2$, the state gradually crosses over to the mixed-valence singlet state.

\subsection{Singlet Phase: Mixed-Valence Singlet}
\label{paper::sec:4.4}

The mixed-valence singlet phase, $|\Delta_d/U-1/2|\lesssim\Gamma_F/U\lesssim 1$, is roughly similar to the mixed-valence regime of the conventional Anderson impurity model. Recall that in the equivalent model~\eqref{paper::eq:8}, the impurity energy level is given by $\epsilon_f=\Delta_d-U/2$ and according to the above phase boundary, $\epsilon_f<\Gamma_F$, and hence the name mixed-valence singlet state.

The most noticeable feature of the mixed-valence singlet region is the emergence of the two energy scales $\Gamma_\pm$ in the local spectral densities, as demonstrated in Fig.~\ref{fig:mixedvalence}. To understand it, we first note that
in this phase ($\Gamma_F>\epsilon_f$) the charge fluctuation on the QD is huge
and at the zeroth order the effects of the on-site interaction $U$ may be ignored.
In the non-interacting picture, the dot Green's functions given by
\begin{subequations}
\begin{align}
G_\up^R(\omega)
& = \frac{1}{\Gamma_+-\Gamma_-}
\left[
\frac{\Gamma_+}{\omega+i\Gamma_+} -
\frac{\Gamma_-}{\omega+i\Gamma_-}
\right] , \\
G_\down^R(\omega)
& = \frac{1}{\Gamma_+-\Gamma_-}
\left[
\frac{\Gamma_+}{\omega+i\Gamma_-} -
\frac{\Gamma_-}{\omega+i\Gamma_+}
\right] ,
\end{align}
\end{subequations}
clearly exhibits two energy scales
\begin{equation}
\Gamma_\pm =
\Gamma_F/2 \pm \sqrt{(\Gamma_F/2)^2 - \epsilon_f^2} \,,
\end{equation}
which represent the relaxation rates predominantly
via the normal tunneling ($c_{k\up}^\dag f_\Up$) and the pair tunneling ($c_{k\up}^\dag f_\Down^\dag$), respectively.
The normal- and pair-tunneling processes are accompanied by phase shift $\pi$ relative to each other and lead to destructive interference; recall  $d_\up=(f_\Up+f_\Down^\dag)/\sqrt2$ from the transformation~\eqref{paper::eq:5}.
The destructive interference is maximal at zero frequency so
that $A_\up(\omega)$ has a dip with a width $\Gamma_-$ inside the central peak
whose width is $\Gamma_+$. For spin $\down$, two processes simply add up so
that two peaks are superposed, displaying a very sharp peak of the width
$\Gamma_-$.

While the non-interacting theory explains the feature of the spectral densities
qualitatively, the NRG results in Section~\ref{paper::sec:3.3} uncover that the interaction $U$ significantly renormalizes $\epsilon_f$ and hence $\Gamma_\pm$
such that $\Gamma_- \ll \Gamma_+ \ll \Gamma_F$.
Especially, $\Gamma_-$ decreases exponentially with decreasing
$\Gamma_F$ and vanishes at the transition point.  One way to investigate such
renormalization effects is again to use the extended variational method in
Section~\ref{paper::sec:4.1} including all
orders~\cite{Varma76a,Gunnarsson83a}. It is, however, out of the scope of the present
work and leave it open for future studies.

\subsection{Singlet Phase: Kondo Singlet}
\label{paper::sec:4.5}

Now we turn to the Kondo singlet regime with $\Gamma_F/U\gg 1, \Delta_d/U$.
% singlet phase where the ground state has $N_S = 0$. Here we focus on the
% S$_\mathrm{K}$ regime where $\Gamma_F$ is the largest energy scale.
In Section~\ref{paper::sec:2.1} we have seen that our model, \eqref{eq:H} or \eqref{paper::eq:8}, is equivalent to the resonant two-level model with negative interaction, \eqref{eq:HSQD:TLM}.
In a recent work \cite{Zitko2011nov} along a different context, it has been found that the resonant two-level model in the large $\Gamma_F$ limit can be bosonized and thus mapped to the anisotropic Kondo model. Interestingly, it was also shown to be related to a quantum impurity coupled to helical Majorana edge modes formed around a two-dimensional topological superconductor. Here we adopt their result to our context, referring the details of the derivation to Ref.~\cite{Zitko2011nov}.

Following the bosonization procedure \cite{Zitko2011nov}, the interacting resonant two-level model is mapped to a bosonized form of the anisotropic Kondo model
\begin{equation}
\label{eq:KM}
H_K
=
\sum_{k\sigma} \epsilon_k c_{k\sigma}^\dag c_{k\sigma}
+ \frac{J_\perp}{2} (S^+ s^- + S^- s^+) + J_z S^z s^z
\end{equation}
with the conduction-band spin $\bfs$ and the impurity spin $\bfS$. Here the
Kondo couplings are identified as
\begin{equation}
J_\perp
= \sqrt8\Delta_d
\end{equation}
and
\begin{equation}
\sqrt2
\left[1 - \frac{2}{\pi} \tan^{-1}\frac{\pi\rho J_z}{4}\right]
= \gamma
:= 1 + \frac{2}{\pi} \tan^{-1}\frac{U}{\Gamma_F}.
\end{equation}
For sufficiently large $\Gamma_F$ compared to $U$, this Kondo model is antiferromagnetic $(J_\perp, J_z >0)$, and the effective Kondo temperature associated with the screening of the magnetic moment is, from the known results on the Kondo model,
\begin{equation}
\label{eq:tk:boson}
T_K^\mathrm{boson}
\sim \frac{\Gamma_F}{2} \left(\frac{2\Delta_d}{\Gamma_F}\right)^{\frac{2}{2-\gamma^2}}.
\end{equation}
As clear from the bosonization procedure, the anisotropic Kondo model
essentially corresponds to the so-called `charge Kondo effect' with the excess
charge on the QD playing the role of the pseudo-spin
\cite{Matveev91a,Iftikhar15a}. More specifically, the charging of $d_\down$
level is mapped onto the pseudo-spin of the Kondo impurity.
Considering that the ferromagnetic lead in our original model has only a single
spin component, this Kondo model should be defined in particle-hole isospin
space of both the dot and the lead. Then, the spin-flip scattering in the
effective Kondo model can be interpreted as the particle-hole scattering in our
original model. For example, the injected particle in the lead is scattered
into the hole, accompanying the inversion of the occupation of $d_\down$
level. Since the change in the occupation of $d_\down$ level is only
possible via the pair tunneling to the superconducting lead, the Kondo
correlation implies that the currents in the ferromagnetic and
superconducting leads are highly correlated.

Here it should be noted that the interpretation based on the bosonization is valid only in the large-$\Gamma_F$ limit because the bosonization procedure requires the unbounded momentum (or dispersion) of a continuum band (whose band width is $\Gamma_F$ in our case) which is to be bosonized. Hence, the mapping to the anisotropic Kondo model cannot be justified in general; in this respect our parameter regime and interpretation are different from those of Ref.~\cite{Zitko2011nov}, where the singlet and doublet phases and the phase transition between them are explained in terms of the effective Kondo model.
One evidence supporting the limitation of the bosonization may come from the comparison between the width of the central peak of $A_\down(\omega)$, which is $T_K$ in the S$_\mathrm{K}$ regime, and the effective Kondo temperature, \eqnref{eq:tk:boson}, predicted from the bosonization [see \figref{fig:Kondo}~(d)]. Two energy scales are in good agreement with each other for $\Gamma_F/U > 1$, as expected. However, for $\Gamma_F/U \lesssim 1$, there is a big discrepancy between them. In addition, the expression~\eqref{eq:tk:boson} fails close to the transition point. It indicates that the region of the singlet phase with small $\Gamma_F$ is not of the Kondo state but of the mixed-valence state, as discussed in the previous section.

\section{Possible Experiments}
\label{sec:experiments}

Up to now, we have elucidated the physical nature of the two phases and, in
particular, classified the different regimes in the singlet phase, mostly based
on the dot spectral densities. One remaining question is how to make a
distinction between the different regimes in experiment. Here we suggest three possible experimental observations: the spin-selective tunneling
microscopy, the current correlation between leads, and the the dynamical
response with respect to the ac gate voltage.

The characteristics of different phases and regimes are well reflected in the spin-dependent spectral density which can be measured by the spin-selective tunneling microscopy applied directly to the quantum dot. It corresponds to adding of an additional ferromagnetic lead very weakly connected to the quantum dot and measuring the differential conductance through it. By altering the polarization of the auxiliary ferromagnetic lead, one can measure the spectral density of the quantum dot for each spin, identifying different phases based on it.

Secondly, as explained in \secref{paper::sec:4.5}, the Kondo scattering in the S$_\mathrm{K}$ regime correlates the currents in the ferromagnetic and superconducting leads, resulting in nontrivial cross-current correlation which can be measured in experiment. Obviously, the average current from the fully polarized ferromagnetic lead to the superconducting lead is still zero in the presence of interacting quantum dot because there is no influx of spin-$\down$ electron from the ferromagnetic lead. However, different from previous works on similar systems \cite{deJong1995feb,Cao2004dec}, the strong interaction in our system makes the currents correlated, though they are zero on average.
Surely, this cross-current correlation should appear in the other regimes of the singlet phase. It can be inferred from the fact that they are divided by crossovers not by sharp transition and that they feature similar spectral densities.  However, in the S$_\mathrm{K}$ regime the current correlation is maximized by the enhanced particle-hole scattering due to the Kondo correlation. Therefore, we expect that the amplitude of the current correlation increases and saturates as one moves toward the S$_\mathrm{K}$ regime.
Experimentally, the current correlation is measured under finite bias because the dc current correlation strictly vanishes at zero bias and the equilibrium low-frequency feature of the correlation is hard to measure in experiment due to decoherence effect. The calculation of the current correlation at large bias is beyond the scope of this work, so we have described this method only qualitatively.

\begin{figure}[!t]
\centering
\includegraphics[width=0.8\columnwidth]{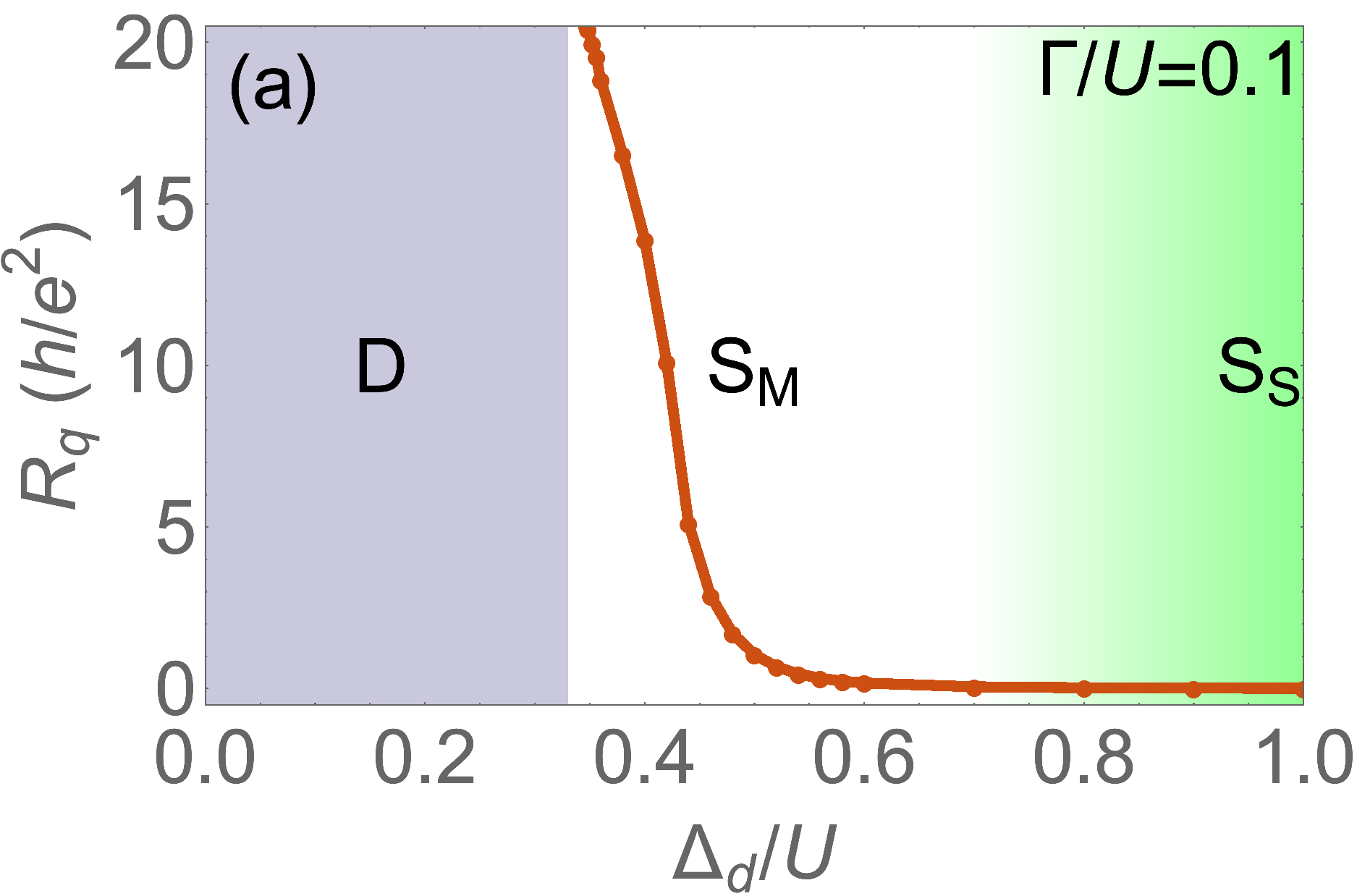}
\includegraphics[width=0.8\columnwidth]{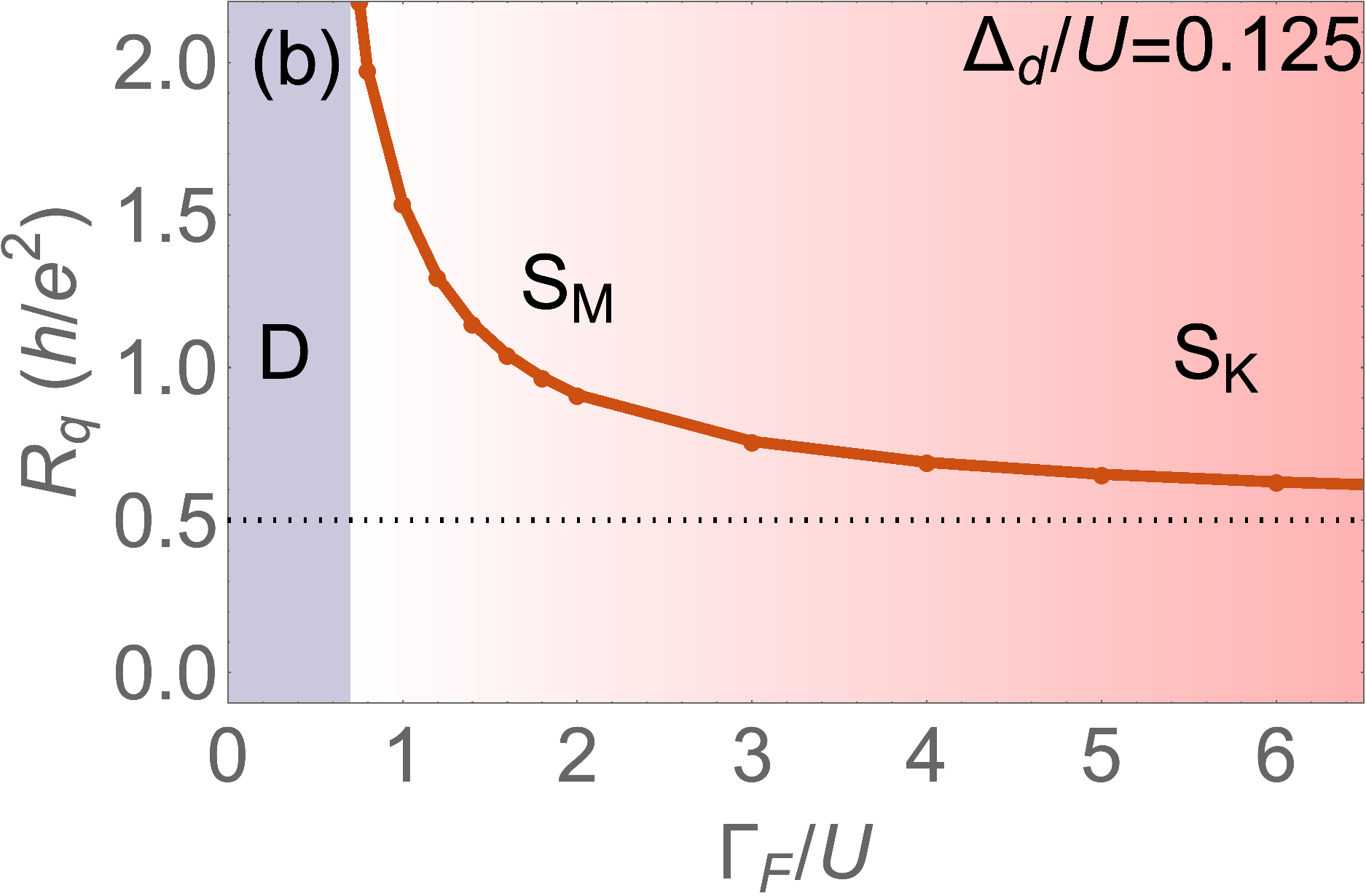}
\caption{Zero-frequency relaxation resistance $R_q(\omega\to0)$ (a) as
  functions of $\Delta_d$ at $\Gamma_F/U=0.1$ along the $bb'$ line in
  \figref{fig:pd} and (b) as functions of $\Gamma_F/U$ at $\Delta_d/U=0.125$
  along the $aa'$ line in \figref{fig:pd}.}
\label{fig:rq}
\end{figure}

The third experimental proposal, which is expected to identify all the phases
and regimes unambiguously, is to measure the charge relaxation resistance in
the zero-frequency limit (a current response to an ac gate
voltage). \Figref{fig:rq} shows the dependence of the zero-frequency
relaxation resistance $R_q(\omega\to0)$ on $\Delta_d$ and $\Gamma_F$.
First, it diverges in the spin doublet regime. Physically, the relaxation resistance is related to the dissipation via the charge
relaxation process of the particle-hole pairs in the lead
\cite{LeeMC2011may}. In the doublet regime, the spin $\down$ level in the dot
is effectively decoupled from the other system and is on resonance, which is
the reason for the two-fold degeneracy \cite{Zitko2011nov}. This resonance
condition enhances the generation of the particle-hole pairs greatly (or
indefinitely in the perturbative sense) \cite{LeeMC2014aug}, leading to
diverging value.

To the contrary, the resistance vanishes in the S$_\mathrm{S}$ regime. In the
presence of the superconductivity, the particle-hole pairs can be generated via
two processes: one is the charge-conserving type ($c_{k\up}^\dag f_\Up$ in
\eqnref{paper::eq:8}) and the other is the pair-tunneling type
($c_{k\up}^\dag f_\Down^\dag$). The particle-hole pair amplitudes of the two
processes are opposite in sign due to the fermion ordering
\cite{LeeMC2014aug}. Also, the cancellation is exact in the zero-frequency
limit of the particle-hole pairs because the weights from the intermediate
virtual states are same for two processes in this limit.
On the other hand, $R_q$ is observed to saturate toward $h/2e^2$ in the
S$_\mathrm{K}$ regime. For a single-channel Fermi-liquid system, the relaxation
resistance is known to have the universal value $h/2e^2$
\cite{Buttiker1993jun,Buttiker1993sep}, and for the conventional Anderson
impurity model in the Kondo regime the resistance becomes $h/4e^2$ since there
are two spin channels which behave like a composite of two parallel resistors
of resistance $h/2e^2$ \cite{LeeMC2011may}. While our system features the
  Kondo correlation in this regime, the resistance is $h/2e^2$ because there is
  only a single channel to generate the particle-hole pairs.
Finally, in the S$_\mathrm{M}$ regime, $R_q$ is finite but strongly depends on
the values of the parameters: it changes continuously from $R_q=\infty$ to the
saturation values, as seen in \figref{fig:rq}. It is known that
\cite{Buttiker1993jun,Buttiker1993sep} the small mesoscopic RC circuit with a
single channel should have a universal value $R_q = h/2e^2$ at zero
temperature as long as it is in the Fermi-liquid state. Non-universal value
of $R_q$ in the S$_\mathrm{M}$, therefore, indicates that the system is in
non-Fermi-liquid states, which makes it distinctive from the S$_\mathrm{K}$
regime. The microscopic origin of the non-universal value of $R_q$ is
explained by the fact that the two opposite effects discussed above are
partially operative simultaneously: the enhancement of the particle-hole
generation due to the high density of states of spin-$\down$ at the Fermi level
(near the spin doublet phase) and the cancellation between the
charge-conserving and pairing processes (near the S$_\mathrm{S}$ regime). The
relative strength of the two effects surely depends on the value of the
parameters.

\section{Conclusion}
\label{sec:conclusion}

Using the NRG method, we have studied the triad interplay of
superconductivity, ferromagnetism, and Kondo effect all together in a QD
coupled to both a superconducting and spin-polarized electrodes as shown
schematically in Fig.~\ref{fig:system} (a).
We have found that unlike the pairwise competition among the three effects,
the triad interplay is ``cooperative'' and leads to a mixed-valence quantum
phase transition between doublet and singlet states.  The singlet phase is in
many respects similar to the mixed-valence state, but connected adiabatically
through crossover either to the superconducting state in the limit of strong
coupling to the superconductor or to the charge Kondo state in the limit of
strong coupling to the spin-polarized lead.
Physical explanations and interpretations based on analytic methods such as
bosonization, scaling theory, and variational method have been provided.
Finally, we have proposed the experimental methods such as the spin-selective
tunneling microscopy, measurement of the cross-current correlation and the
charge relaxation resistance in order to distinguish the different phases and
regimes.

Even though our study has found out the key characteristics of the
ferromagnet-quantum dot-superconductor system, it still leaves much room for
further studies.
First, one can lift the particle-hole symmetry condition used in this
work. Then, due to the ferromagnetic proximity effect, it induces an
effective Zeeman splitting (or exchange field), which would form subgap
states in the dot. Moreover, the breaking of the particle-hole symmetry for
spin-$\down$ level is expected to induce an effective Zeeman field for the
Kondo model in the S$_\mathrm{K}$ regime, shifting the
phase boundaries \cite{endnote:1}.
Secondly, the strong superconductivity condition
($\Delta_0 \gg U, \Gamma_S, \Gamma_F$) also can be lifted so that the spin
Kondo-dominated state ($T_K > \Delta_0$) can arise. Then, the S$_\mathrm{S}$
regime will be replaced by the Kondo state. In this case, one may observe the
interesting crossover from the spin Kondo state to the charge Kondo state.
Finally, the study can go beyond the equilibrium case by applying a finite
bias which is still below the superconducting gap. As discussed in
\secref{sec:experiments}, the calculation of the cross-current correlation at
finite bias is important for experimental verification. Although the
non-equilibrium condition in the presence of a strong interaction is
challenging, it is worth doing in the experimental point of view.

\section*{Acknowledgments}

This work was supported by the the National Research
Foundation (Grant Nos. 2011-0030046, NRF-2017R1E1A1A03070681, and 2018R1A4A1024157)
and the Ministry of Education (through the BK21 Plus Project) of Korea.

\bigskip\bigskip

\appendix

\section{Variational Method for the Single-Impurity Anderson Model}
\label{sec:vmsiam}

Here we apply the variational method to the conventional Anderson impurity model described by
\begin{multline}
H_\mathrm{A}
=
\epsilon_d \sum_\sigma n_\sigma + U n_+ n_-
+ \sum_{k\sigma} \epsilon_k \epsilon_{k\sigma}^\dag \epsilon_{k\sigma} \\{}
+ \sum_{k\sigma} \left(t_\sigma d_\sigma^\dag c_{k\sigma} + (h.c.)\right)
\end{multline}
in the same way in the text. The conventional Anderson impurity model is different from our model in two
points: one is that the lead has two (spin) channels and the second is that the
tunneling conserves the spin. The ansatz for spin singlet and doublet states
constructed in the similar way as in \eqnref{eq:ansatz} is
\begin{widetext}
\begin{subequations}
\label{eq:ansatz:siam}
\begin{align}
\ket{S}
& =
\left(
\alpha_0 + \sum_{k<k_F,\sigma} \alpha_{k\sigma} d_\sigma^\dag c_{k\sigma}
+ \sum_{k<k_F,k'>k_F,\sigma} \alpha_{kk'\sigma} c_{k'\sigma}^\dag c_{k\sigma}
\right)
\ket{\mathrm{FS}}_0
\\
\ket{D_\sigma}
& =
\left(
\beta_{0\sigma} d_\sigma^\dag + \sum_{k>k_F} \beta_{k\sigma} c_{k\sigma}^\dag
+
\sum_{k>k_F,k'<k_F,\sigma'} \beta_{kk'\sigma'\sigma}
c_{k\sigma}^\dag d_{\sigma'}^\dag c_{k'\sigma'}
\right)
\ket{\mathrm{FS}}_0.
\end{align}
\end{subequations}
\end{widetext}
Now we minimize the energy expectation value with respect to these states by
applying the Lagrange multiplier method under the normalization constraint
$\Braket{S|S} = \Braket{D_\sigma|D_\sigma} = 1$. Then, one can obtain the
following coupled differential equations:
\begin{subequations}
\begin{align}
\epsilon_S \alpha_0
& =
- \epsilon_d \alpha_0
+ \sum_{k<k_F,\sigma} t_\sigma^* \alpha_{k\sigma}
\\
\epsilon_S \alpha_{k\sigma}
& =
t_\sigma \alpha_0 - \epsilon_k \alpha_{k\sigma}
+ t_\sigma \sum_{k'>k_F} \alpha_{kk'\sigma}
\\
\epsilon_S \alpha_{kk'\sigma}
& =
t_\sigma^* \alpha_{k\sigma}
+ (\epsilon_{k'} - \epsilon_k - \epsilon_d) \alpha_{kk'\sigma}
\end{align}
\end{subequations}
and
\begin{subequations}
\begin{align}
\epsilon_D \beta_{0\sigma}
& = t_\sigma \sum_{k>k_F} \beta_{k\sigma}
\\
\epsilon_D \beta_{k\sigma}
& =
t_\sigma^* \beta_{0\sigma} + (\epsilon_k - \epsilon_d) \beta_{k\sigma}
+ \sum_{k'<k_F,\sigma'} t_{\sigma'}^* \beta_{kk'\sigma'\sigma}
\\
\epsilon_D \beta_{kk'\sigma'\sigma}
& = t_{\sigma'} \beta_{k\sigma} + (\epsilon_k - \epsilon_{k'}) \beta_{kk'\sigma'\sigma}.
\end{align}
\end{subequations}
Up to the first order (by setting $\alpha_{kk'\sigma} =
\beta_{kk'\sigma'\sigma} = 0$), the closed-form equations for $\epsilon_S$ and
$\epsilon_D$ are given by
\begin{subequations}
\label{eq:vm:siam}
\begin{align}
\epsilon_S
& =
- \epsilon_d
- \frac{\Gamma_++\Gamma_-}{\pi}
\ln\left(1 + \frac{D}{|\epsilon_S|}\right)
\\
\epsilon_D
& =
-
\frac{\Gamma_\mu}{\pi}
\ln\left(1 + \frac{D}{|\epsilon_D| - \epsilon_d}\right)
\end{align}
\end{subequations}
which is basically same as \eqnref{eq:vm1} except the fact that the dot-lead
hybridization is increased since the conventional Anderson impurity model has two spin channels in the lead.
Up to the second order, the self-consistent equations for $\epsilon_S$ and
$\epsilon_D$ read
\begin{subequations}
\begin{align}
\epsilon_S
& =
- \epsilon_d
-
\sum_\sigma \frac{\Gamma_\sigma}{\pi}
\int_0^D \frac{d\epsilon'}%
{\displaystyle
  \epsilon' + |\epsilon_S|
  -
  \frac{\Gamma_\sigma}{\pi}
  \ln\frac{\epsilon' + D + |\epsilon_S| + |\epsilon_d|}%
  {\epsilon' + |\epsilon_S| + |\epsilon_d|}}
\\
\epsilon_D
& =
-
\frac{\Gamma_\sigma}{\pi}
\int_0^D \frac{d\epsilon'}%
{\displaystyle
  \epsilon' + |\epsilon_{D\sigma}| + |\epsilon_d|
  -
  \frac{\Gamma_+ + \Gamma_-}{\pi}
  \ln\frac{\epsilon' + D + |\epsilon_{D\sigma}|}{|\epsilon_d| + |\epsilon_{D\sigma}|}}.
\end{align}
\end{subequations}

\bigskip

%%%% References

\bibliographystyle{apsrev}
\bibliography{Paper}

\begin{thebibliography}{60}
\bibitem[{end({\natexlab{a}})}]{endnote:2}
\bibinfo{note}{For finite polarization, in general the effects of
  contact-induced exchange field should be considered; see
  \cite{Sahoo05a,ChoiMS06d,ChoiMS06g,Wojcik14a,Weymann15a}. As long as the
  exchange Zeeman splitting is sufficiently larger than the Kondo temperature,
  however, they are negligible and our main results are not affected
  qualitatively even for finite spin polarization.}
\end{thebibliography}

\end{document}